\documentclass{emulateapj}
\usepackage{psfig,enumerate}
\usepackage{xcolor}
\usepackage{color}
\usepackage{atbegshi}
\usepackage{url}
\usepackage[caption=false]{subfig}
\usepackage[pdfpagemode={UseOutlines},bookmarks=true,bookmarksopen=true,
   bookmarksopenlevel=0,bookmarksnumbered=true,hypertexnames=false,
   colorlinks,linkcolor={blue},citecolor={blue},urlcolor={red},
   pdfstartview={FitV},unicode,breaklinks=true]{hyperref}
\def\hackaltaffiltext#1#2{\AtBeginShipoutNext{\footnotetext[#1]{#2}\stepcounter{footnote}}}

\def\apjl{ApJL}

\def\apjs{ApJS}

\def\aap{{A\&A}}

\begin{document}

\title{Exploring the stellar age distribution of the Milky Way Bulge using APOGEE}

\shorttitle{APOGEE Bulge Ages}
\shortauthors{Hasselquist et al.}

\author{
Sten Hasselquist\altaffilmark{1,$\dagger$},
Gail Zasowski\altaffilmark{1},
Diane K. Feuillet\altaffilmark{2},
Mathias Schultheis\altaffilmark{3},
David M. Nataf\altaffilmark{4}, 
Borja Anguiano\altaffilmark{5},
Rachael L. Beaton\altaffilmark{6,7},
Timothy C. Beers\altaffilmark{8},
Roger E. Cohen\altaffilmark{9},
Katia Cunha\altaffilmark{10,11},
Jos{\'e} G.\ Fern{\'a}ndez-Trincado\altaffilmark{12},
D. A. Garc{\'i}a-Hern{\'a}ndez\altaffilmark{13,14},
Doug Geisler\altaffilmark{15,16,17},
Jon A. Holtzman\altaffilmark{18},
Jennifer Johnson\altaffilmark{19},
Richard R. Lane\altaffilmark{12,20},
Steven R. Majewski\altaffilmark{5},
Christian Moni Bidin\altaffilmark{21},
Christian Nitschelm\altaffilmark{22},
Alexandre Roman-Lopes\altaffilmark{17},
Ricardo Schiavon\altaffilmark{23},
Verne V. Smith\altaffilmark{24},
Jennifer Sobeck\altaffilmark{25}
}

\altaffiltext{1}{Department of Physics \& Astronomy, University of Utah, Salt Lake City, UT, 84112, USA (stenhasselquist@astro.utah.edu)}
\altaffiltext{$\dagger$}{NSF Astronomy and Astrophysics Postdoctoral Fellow}
\altaffiltext{2}{Lund Observatory, Department of Astronomy and Theoretical Physics, Box 43, SE-221\,00 Lund, Sweden}
\altaffiltext{3}{Laboratoire Lagrange, Universit\'{e} C\^{o}te d'Azur, Observatoire de la C\^{o}te d'Azur, 06304, Nice, France}
\altaffiltext{4}{Center for Astrophysical Sciences and Department of Physics and Astronomy, The Johns Hopkins University, Baltimore, MD 21218}
\altaffiltext{5}{Department of Astronomy, University of Virginia, Charlottesville, VA, 22904, USA}
\altaffiltext{6}{Department of Astrophysical Sciences, Princeton University, 4 Ivy Lane, Princeton, NJ 08544, USA}
\hackaltaffiltext{7}{The Observatories of the Carnegie Institution for Science, 813 Santa Barbara St., Pasadena, CA 91101, USA}
\altaffiltext{8}{Department of Physics, University of Notre Dame, \& JINA Center for the Evolution of the Elements, Notre Dame, IN, 46556 USA}
\altaffiltext{9}{Space Telescope Science Institute, 3700 San Martin Drive, Baltimore, MD 21218, USA}
\altaffiltext{10}{Steward Observatory, The University of Arizona, Tucson, AZ, 85719, USA}
\altaffiltext{11}{Observat\'{o}rio Nacional, 20921-400 So Crist\'{o}vao, Rio de Janeiro, RJ, Brazil}
\altaffiltext{12}{Instituto de Astronom\'ia y Ciencias Planetarias, Universidad de Atacama, Copayapu 485, Copiap\'o, Chile}
\altaffiltext{13}{Instituto de Astrofísica de Canarias (IAC), E-38205 La Laguna, Tenerife, Spain}
\altaffiltext{14}{Universidad de La Laguna (ULL), Departamento de Astrof{\'i}sica, E-38206 La Laguna, Tenerife, Spain}
\altaffiltext{15}{Departamento de Astronom{\'i}a, Casilla 160-C, Universidad de Concepcion, Chile}
\altaffiltext{16}{Instituto de Investigación Multidisciplinario en Ciencia y Tecnología, Universidad de La
Serena. Avenida Raúl Bitrán S/N, La Serena, Chile}
\altaffiltext{17}{Departamento de Física y Astronom{\'i}a, Facultad de Ciencias, Universidad de La Serena. Av. Juan Cisternas 1200, La Serena, Chile}
\altaffiltext{18}{Department of Astronomy, New Mexico State University, Las Cruces, NM 88003, USA}
\altaffiltext{19}{Department of Astronomy, The Ohio State University, 140 W. 18th Ave., Columbus, OH 43210, USA}
\altaffiltext{20}{Instituto de Astrof{\'i}sica, Pontificia Universidad Cat\'{o}lica de Chile, Av. Vicuna Mackenna 4860, 782-0436 Macul, Santiago, Chile}
\altaffiltext{21}{Instituto de Astronom{\'i}a, Universidad Cat\'{o}lica del Norte, Av. Angamos 0610, Antofagasta, Chile}
\altaffiltext{22}{Centro de Astronom{\'i}a (CITEVA), Universidad de Antofagasta, Avenida Angamos 601, Antofagasta 1270300, Chile}
\altaffiltext{23}{Astrophysics Research Institute, Liverpool John Moores University, IC2, Liverpool Science Park, 146 Brownlow Hill, Liverpool L3 5RF, UK}
\altaffiltext{24}{National Optical Astronomy Observatory, 950 North Cherry Ave, Tucson, AZ 85719}
\altaffiltext{25}{Department of Astronomy, University of Washington, Seattle, WA, 98195, USA}

\begin{abstract}

We present stellar age distributions of the Milky Way (MW) bulge region using ages for $\sim$6,000 high-luminosity ($\log(g) < 2.0$), metal-rich ($\rm [Fe/H] \ge -0.5$) bulge stars observed by the Apache Point Observatory Galactic Evolution Experiment (APOGEE). Ages are derived using {\it The Cannon} label-transfer method, trained on a sample of nearby luminous giants with precise parallaxes for which we obtain ages using a Bayesian isochrone-matching technique. We find that the metal-rich bulge is predominantly composed of old stars ($>$8 Gyr). We find evidence that the planar region of the bulge ($|Z_{\rm GC}| \le 0.25$ kpc) enriched in metallicity, $Z$, at a faster rate ($dZ/dt \sim$ 0.0034 ${\rm Gyr^{-1}}$) than regions farther from the plane ($dZ/dt \sim$ 0.0013 ${\rm Gyr^{-1}}$ at $|Z_{\rm GC}| > 1.00$ kpc). We identify a non-negligible fraction of younger stars (age $\sim$ 2--5 Gyr) at metallicities of $\rm +0.2 < [Fe/H] < +0.4$. These stars are preferentially found in the plane ($|Z_{\rm GC}| \le 0.25$ kpc) and between $R_{\rm cy} \approx 2-3$ kpc, with kinematics that are more consistent with rotation than are the kinematics of older stars at the same metallicities. We do not measure a significant age difference between stars found in and outside of the bar. These findings show that the bulge experienced an initial starburst that was more intense close to the plane than far from the plane. Then, star formation continued at super-solar metallicities in a thin disk at 2 kpc $\lesssim R_{\rm cy} \lesssim$ 3 kpc until $\sim$2 Gyr ago.

\end{abstract}

\section{Introduction}

The vast majority of disk galaxies in the local Universe harbor an over-density of light in their centers, commonly referred to as a ``bulge''. These bulges appear with a variety of structures and stellar populations, presumably resulting from different evolutionary processes; spheroidal ``classical'' bulges dominate in the most massive disk galaxies, while pseudobulges (mostly bars) are more common in Milky Way (MW)-mass galaxies \citep[e.g.,][]{Fisher_2011_bulgedemographics}.  When these bulges --- barred or not --- form, and how they evolve over time, are still open questions; likely, these complex systems grew through some combination of accretion of stars that now reside in the center of the MW \citep[e.g.,][] {Tumlinson2010} and in situ star formation in a disk that later buckled into the structures observed today \citep[see review in][]{Athanassoula2005}.

The MW itself has a central, asymmetric, boxy overdensity of light first measured in integrated infrared (IR) photometry \citep[e.g.,][] {Weiland1994,Dwek1995}. Additional photometric and kinematic studies soon revealed that this boxy, peanut-shaped structure is likely a result of seeing a bar structure edge-on \citep[e.g.,][]{Hammersley1994,Athanassoula2005,McWilliam&Zoccali2010,Nataf2010,Wegg&Gerhard2013}, and that the inner MW harbors a barred mass distribution with a semi-major axis $\simeq$ 5 kpc (e.g., \citealt{Wegg2015,Bovy2019}) and major-minor axis ratio of 0.4 \citep{Bovy2019}.
This barred central structure, which we will refer to in this work as the ``bulge'', contains $\sim$50\% of the MW's stellar mass \citep[]{Licquia2015}, with $\sim$60\% of that bulge mass residing in the bar structure and 40\% in the inner disk \citep[e.g.,][]{Portail2017}. 
In order to understand the full picture of how our MW Galaxy formed and evolved, we must understand the star-formation and chemical-enrichment histories of this critical region that contains the majority of stellar mass (e.g., \citealt{Rich2013,McWilliam2016,Barbuy2018}).

The stars that reside in the bulge span $\sim$2.5 dex in metallicity (nearly the full range observed across the MW); the metallicity distribution function (MDF) changes dramatically as a function of position in the inner Galaxy, resulting in an average negative metallicity gradient with height above the MW plane \citep[e.g.,][]{Zoccali2008,Johnson2013,Rojas-Arriagada2014,Zoccali2017,Zoccali2018,GarciaPerez_2018_dr12mdf,Fragkoudi2018}. In addition, the kinematics of the stellar populations are correlated with their chemistry, with the lower-metallicity stars on more spheroidal orbits and the higher-metallicity stars on more ``bar-like'' orbits (e.g., \citealt{Hill2011,Ness2013,Ness2016a,Zasowski2016,Barbuy2018}). Despite this spatially variant, broad MDF, the detailed elemental abundances of the bulge stars appear to be not only relatively homogeneous throughout the bulge, but also nearly identical to the chemical abundance pattern of thick-disk stars at the solar radius (e.g., \citealt{Rojas-Arriagada2017,Haywood2018,Zasowski2019}, but see e.g., \citealt{Johnson2014}). An exception to this similarity arises from the non-negligible fraction of inner Galaxy stars originating from dissolved globular clusters \citep[e.g.,][]{Schiavon2017,Fernandez-Trincado2017}.

These chemodynamical patterns suggest that the bulge formed many of its stars early, in a rapid star-formation event (e.g., \citealt{McWilliam1994,Fulbright2007,Johnson2012}), and probably in a disk that later buckled into a boxy bar, giving rise to the metal-rich stars on radial bar-like orbits (e.g., \citealt{Athanassoula2016}). However, major uncertainties remain regarding the timing of this buckling, the detailed structure of the disk before buckling (e.g., \citealt{Fragkoudi2017}), and the extent to which star formation has proceeded since this buckling event. 

To answer these questions, we need ages for large numbers of bulge stars at {\it all} Galactocentric radii ($R_{\rm cy}$) and distances from the plane ($|Z_{\rm GC}|$), including in the midplane itself.  Most age studies of the bulge to date have shown that all, or nearly all, of the stars appear to be old \citep[$>$~9-10~Gyr; e.g.,][]{Zoccali2003,Clarkson2011,Barbuy2018,Renzini2018}. However, other groups found evidence for significant fractions of bulge stars with ages $<$~8~Gyr \citep[e.g.,][]{vanLoon2003,Bensby2013,Catchpole2016,Bensby2017}, which are expected from some simulations, especially if only stars in the plane are considered \citep[e.g.,][]{Ness2014}. Such a large fraction of intermediate-age stars is consistent with some CMD-based studies \citep[e.g.,][]{Holtzman1993,Haywood2016}, but inconsistent with others \citep[e.g.,][]{Clarkson2009,Gennaro2015,Surot_2019_VVVages1}. We refer the reader to \citet{Nataf2016} and \citet{Barbuy2018} for recent reviews, and note that these statements only apply to the larger bulge population beyond the innermost few hundred parsecs, where there is known to be recent and ongoing star formation (e.g., \citealt{Morris_1996_MWCenter,Longmore_2013_MwSfr})

In addition to the ambiguous presence of relatively young stars in the bulge, the spatial variations of the mean stellar age and the stellar age distribution are highly uncertain. The majority of bulge age studies have been limited to pencil-beam fields, typically not in the high-extinction midplane. Recent work from \citet{Bovy2019} argues that the bar appears distinctly older (and more metal-poor) than the inner disk, with a mean age of $\simeq$8~Gyr. In contrast, \citet{Wegg2019} find evidence for a metal-rich bar population younger than the local disk. We discuss both of these findings in the context of our analysis in \S \ref{sec:results} below.

Large-scale spectroscopic surveys continue to provide an ever-growing amount of chemical, kinematical, and most recently, age information for stars across the MW, allowing us to explore numerous formation scenarios in different parts of the Galaxy. The Apache Point Observatory Galactic Evolution Experiment \citep[APOGEE;][]{Majewski2017} is a survey particularly well-equipped to study the inner MW's evolution. Because APOGEE operates in the near-IR, the survey can observe stars in the MW center, even through the thick dust in the mid-plane of the Galaxy. In addition, the abundance of carbon and nitrogen features in its spectra allow for precise determination of carbon and nitrogen abundances, which in turn can be correlated with asteroseismology derived masses and mapped to stellar ages. 

Numerous studies have explored the MW disk in this way: e.g., \citet{Masseron&Gilmore2015}, \citet{Martig2016b}, \citet{Ness2016b}, \citet{Mackereth2019b}, and \citet{Hasselquist2019a}. Works that map ages onto APOGEE stars have usually relied on the exquisite APOKASC \citep{Pinsonneault2014} and APOKASC-2 \citep{Pinsonneault2018} stellar masses and ages as a training sample. However, the APOGEE stars observed in the bulge are typically much cooler and more luminous than those APOKASC stars with precise masses, meaning that computing ages for bulge stars is impossible without significant extrapolation.

In this paper, we use a new training set to compute ages for $\sim$ 46,000 stars in the MW Galaxy, and analyze the results for $\sim$ 6,000 stars in the bulge, nearly all of which are beyond the parameter space of the APOKASC-2 sample. We describe our data and bulge sample selection in \S\ref{sec:obs}, the spectral age information in \S\ref{sec:age_info}, our training set in \S\ref{sec:train_set}, and our implementation of  {\it The Cannon} in \S\ref{sec:cannon}. Our results --- including mean age maps, the presence of young stars, and on-/off-bar differences --- are presented in \S\ref{sec:results} and discussed in \S\ref{sec:disc}.

\section{Observations and Sample Selection}
\label{sec:obs}

Observations were taken as part of the Apache Point Observatory Galactic Evolution Experiment \citep[APOGEE;][]{Majewski2017}, part of the fourth iteration of the Sloan Digital Sky Survey (SDSS-IV; \citealt{Blanton2017}). The APOGEE instruments are high-resolution, near-infrared spectrographs \citep{Wilson2019} observing from both the Northern Hemisphere at Apache Point Observatory (APO) using the SDSS 2.5m telescope \citep{Gunn2006}, and the Southern Hemisphere at Las Campanas Observatory (LCO) using the  2.5m du Pont telescope \citep{Bowen73}. As of December 2019 the dual APOGEE instruments have observed some 500,000 stars across the MW, targeting these stars with selections described in \citet{Zasowski2013} and \citet{Zasowski2017}, with updates to the targeting plan described in Santana et al. (2020) and Beaton et al. (2020). 

Spectra are reduced as described in \citet{Nidever2015} and analyzed using the APOGEE Stellar Parameters and Chemical Abundance Pipeline (ASPCAP, \citealt{Garcia-Perez2016}). A detailed analysis of the accuracy and precision of the stellar parameters and abundances can be found in \citet{Holtzman2018} and \citet{Jonsson2018}. Our analysis uses results from the 16th Data Release (DR16) of the SDSS collaboration \citep{dr16}, which is the first data release containing data from the Southern instrument. Further explanations and assessments of this data release, including quantification of potential offsets between the Northern and Southern spectrographs, can be found in \citet{Jonsson2020}.

This work focuses on the APOGEE stars that reside in the bulge, which we define using Galactic cylindrical coordinates, $R_{\rm cy}$ = $\sqrt{X_{\rm GC}^{2}+Y_{\rm GC}^{2}}$ and $|Z_{\rm GC}|$. We transform the APOGEE stars into these coordinates based on their position in the sky and distance, after adopting a Solar Galactocentric position of $X_{\rm GC}$ = -8.125 kpc \citep{gravcollab2018} and $Z_{\rm GC}$ = 20.8 pc, as was done in \citet{Bovy2019}. We adopt the stellar distances derived using \texttt{astroNN}\footnote{\url{https://www.sdss.org/dr16/data\_access/value-added-catalogs/?vac\_id=the-astronn-catalog-of-abundances,-distances,-and-ages-for-apogee-dr16-stars} and \url{https://github.com/henrysky/astroNN}} \citep{Leung2019}. These distances have uncertainties $\sim$ 15\% at the location of the bulge, but both \citet{Queiroz2020} and \citet{Bovy2019} noticed that the DR16 \texttt{astroNN} distances are slightly underestimated at locations $>$ 5 kpc from the Sun. Therefore, we apply the same distance correction derived and implemented by \citet{Bovy2019}.

We show the resultant spatial distribution of our stellar sample in both the $X_{\rm GC}-Y_{\rm GC}$ plane and $R_{\rm cy}-Z_{\rm GC}$ plane in the left two columns of Figure \ref{fig:dist}. We define the bulge to be all stars with $R_{\rm cy} < $ 3.5 kpc and $|Z_{\rm GC}| < $ 1.5 kpc. Only stars that meet the following criteria are plotted in Figure \ref{fig:dist}:

\begin{figure*}[t]
\includegraphics[width=1.0\hsize,angle=0]{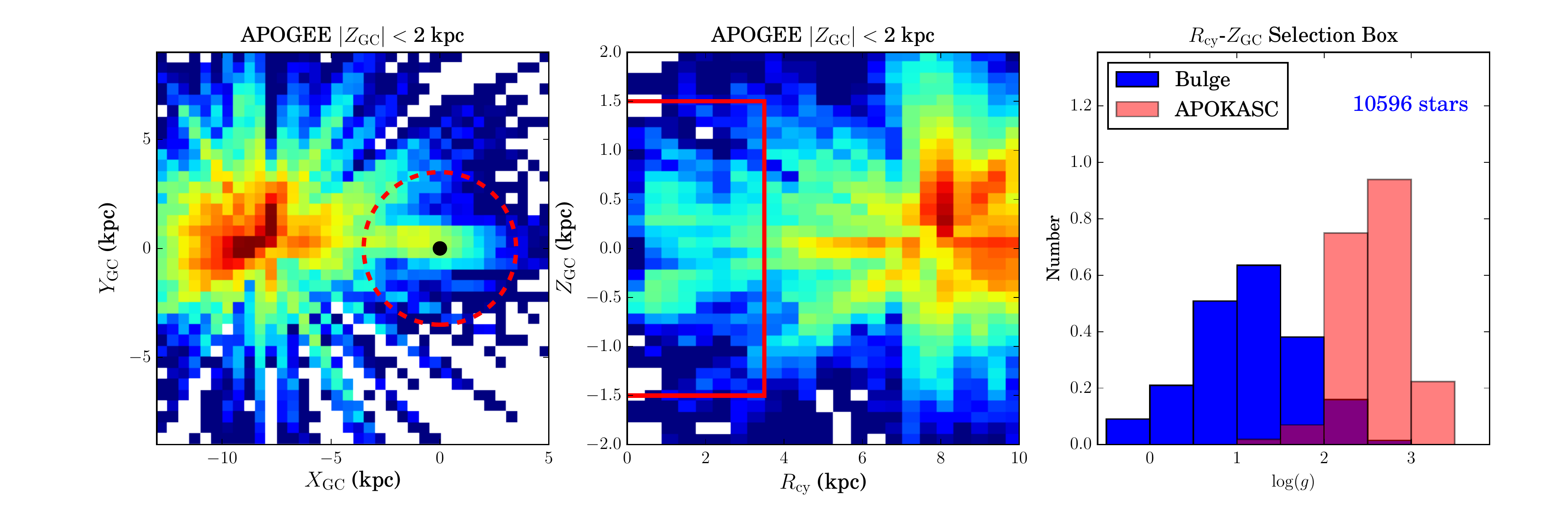}
\caption{The distribution of the APOGEE stellar sample in the $X_{\rm GC}-Y_{\rm GC}$ plane (left), $R_{\rm cy}-Z_{\rm GC}$ plane (middle), and $\log(g)$ distribution for stars in the bulge (right) for the APOGEE sample using the \texttt{astroNN} distances. The right panel also includes the $\log(g)$ distribution of stars in the APOKASC sample with ages from asteroseismic masses. The red circle on the left plot and the red lines on the middle plot show the spatial cuts we use to define our bulge sample, as described in the text.}
\label{fig:dist}
\end{figure*}

\begin{itemize}
    \item S/N $>$ 70 per pixel
    \item ~[Fe/H] $>$ -0.5\footnote{As described more in \S \ref{sec:age_info}, stars below this metallicity no longer have clean age-sensitive features in their spectra.}
    \item No STAR\_BAD bit of ASPCAPBAD flag set
    \item $\log(g)$ $<$ 3.3
\end{itemize}

The third panel of Figure \ref{fig:dist} shows the $\log(g)$ distribution of the stars that fall inside this bulge cut for each distance sample (blue histogram). The APOKASC sample, which has been used in the past as a training set for data-driven age determination methods, is shown in red. As discussed further in \S \ref{sec:ages}, previous works that have employed {\it The Cannon} or similar techniques to derive ages of APOGEE stars use training sets, such as APOKASC \citep{Pinsonneault2014}, that have few stars with $\log(g)$ $<$ 2.0 and almost no stars with $\log(g)$ $<$ 1.5. Therefore, to derive ages for these ``luminous giants'' ($\log(g)$ $<$ 2.0) that make up the bulge APOGEE sample, \emph{we must use a different age-training sample.}

\section{Stellar Ages}
\label{sec:ages}

The goal of this work is to derive ages for the luminous giants ($\log(g)$ $<$ 2.0) that primarily comprise the APOGEE bulge sample. We use {\it The Cannon} \citep{Ness2015} to derive ages, but because of the vastly different $\log(g)$ distributions highlighted in Figure \ref{fig:dist}, we require a new training set to derive ages using this tool. In this section, we confirm there is age information in the APOGEE spectra for the luminous giants (\S \ref{sec:age_info}), describe a new training set we will use to derive labels with {\it The Cannon} (\S \ref{sec:train_set}), and discuss the application of {\it The Cannon} to our data set (\S \ref{sec:cannon}). 

\subsection{Age Information in APOGEE Spectra}
\label{sec:age_info}

The age information in the APOGEE spectra of red giant stars primarily comes from carbon and nitrogen molecular features. This is because the birth [C/N] abundance of a star is affected by first dredge-up as it ascends the red giant branch. This dredge-up operates in such a way that the resultant [C/N] abundance ratio after the star has ascended the red giant branch depends on the mass of the progenitor, such that more massive stars have lower [C/N] abundances (e.g., \citealt{Gratton2000,Martell2008,Salaris2015}). Stellar models can then be invoked to obtain an age for an RGB star of a given mass and metallicity. There are now several works in the literature where this dependence has been exploited to interpret [C/N] abundance variations across the Galaxy as age variations (\citealt{Masseron&Gilmore2015,Hasselquist2019b}), and even works that map ages directly on to stars observed by APOGEE (e.g., \citealt{Martig2016b,Ness2016b,Mackereth2019b}).

\begin{figure*}[t]
\includegraphics[width=1.0\hsize,angle=0]{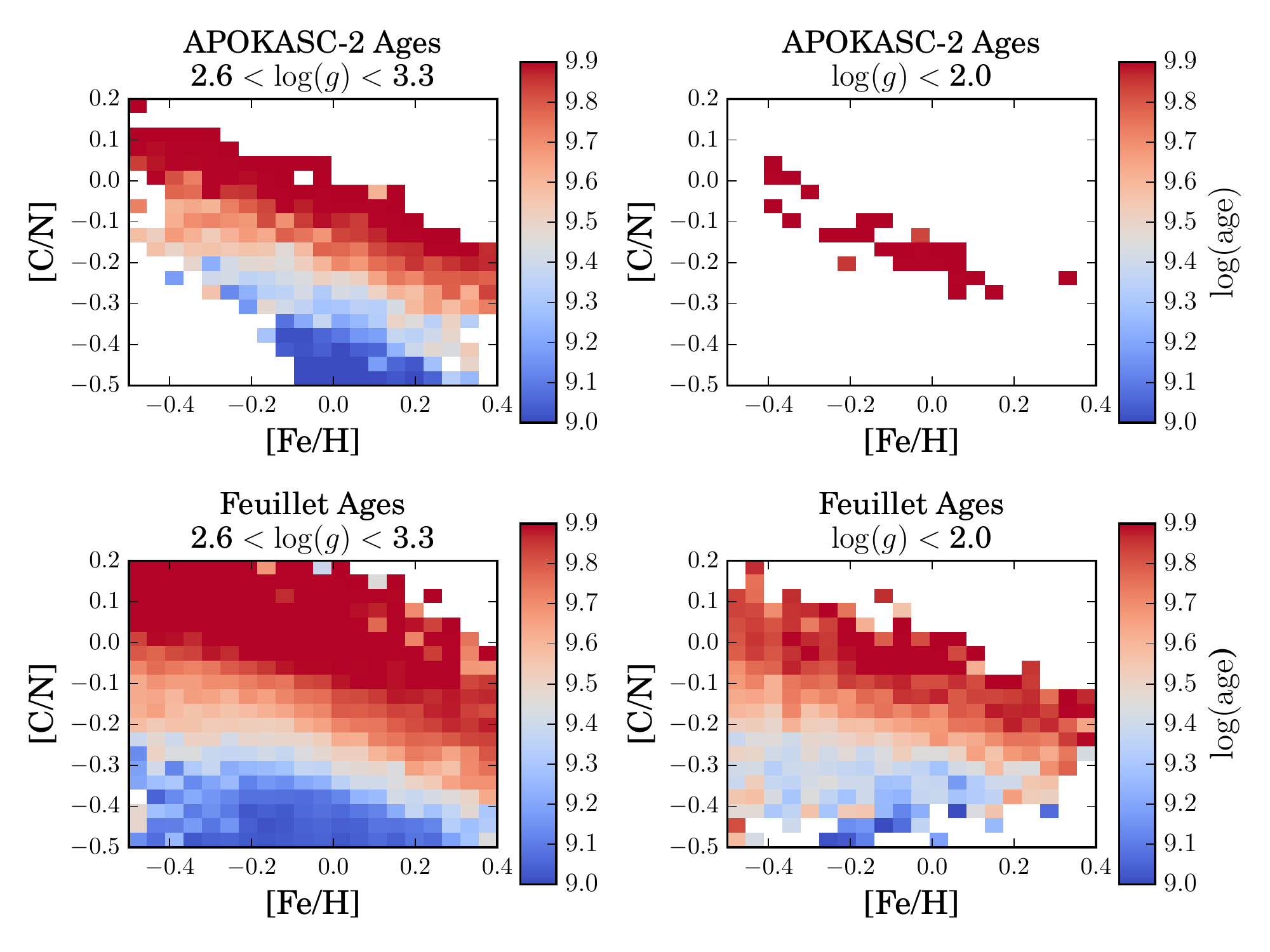}
\caption{The [C/N]-[Fe/H] plane colored by mean {$\rm \log(age)$}. The top row shows this plane for the APOKASC-2 age training sample and the bottom row shows this plane for the ``Feuillet'' ages, described more in \S \ref{sec:train_set}. The left column compares training sets for stars below the red clump and the right column compares training sets for stars above the red clump, which is where the majority of the APOGEE bulge sample studied here are found.} 
\label{fig:apokasc}
\end{figure*}

These studies that derive ages for APOGEE stars all rely on the APOKASC and APOKASC-2 sample \citep{Pinsonneault2014,Pinsonneault2018} as a training set. These stars have precise masses ($\lesssim$ 10\%) derived using asteroseismology. Ages are then inferred using stellar evolutionary models. However, this APOKASC sample is limited in the parameter space covered; specifically, it is lacking in stars with $\log(g)$ $<$ 2.0 and nearly completely devoid of stars with $\log(g)$ $<$ 1.5. Therefore, it is not an ideal training set for deriving ages for the APOGEE bulge stars, the majority of which have $\log(g)$ $<$ 2.0 (as shown in the right panel of Figure \ref{fig:dist}), as it often results in ages derived from model extrapolation. This is summarized in the top row of Figure \ref{fig:apokasc}, where we show that, while the age coverage of the APOKASC-2 sample is quite good in the [C/N]-[Fe/H] plane for stars on the lower giant branch (2.6 $<$ $\log(g)$ $<$ 3.3), there are far fewer stars with $\log(g)$ $<$ 2.0. The stars that are there are old, on average, and do not span the full range of [C/N]-[Fe/H] space covered by the bulge sample, which highlights the need for a different training set.

\subsection{Luminous Star Training Set}
\label{sec:train_set}

To derive bulge ages using {\it The Cannon}, we must find a large sample of high-luminosity ($\log(g)$ $<$ 2.0) stars that have ages. Fortunately, several studies in the literature (e.g., \citealt{Feuillet2016,Anders2018,Feuillet2018,Queiroz2018}) have shown that it is possible to use Bayesian isochrone matching to derive precise ($\sim$ 0.2 dex uncertainty in {$\rm \log(age)$}) ages for red giant stars if the distance is known to better than 10\%. With \emph{Gaia} DR2 \citep{GaiaDR2}, there are now several thousand low-gravity stars in the APOGEE sample with distances more accurate than 10\%. These stars are distributed across the MW disk from 6 kpc $<$ $R_{\rm cy}$ $<$ 11 kpc, resulting in a training set that spans a range of stellar populations. We refer to the ages derived in the methods described below as ``Feuillet Ages''. 

\subsubsection{Deriving Ages for the Training Set}

Briefly, ages for the APOGEE luminous giant training set stars were derived using a simple Bayesian isochrone-matching method described by \citet{Jorgensen2005}, which produces an age probability distribution function (PDF). To derive the PDF for each star, we compute a likelihood function comparing the measured parameters from each star with scaled-solar PARSEC isochrones \citep{Bressan2012} and an assumed prior star-formation history (SFH), metallicity distribution function (MDF), and initial mass function (IMF). The age of each star is assigned to be the age-weighted mean of the final PDF, and the age uncertainty is the formal dispersion in the PDF. 

The measured parameters used for the likelihood function are effective temperature, surface gravity, $\alpha$-adjusted metallicity, and absolute $K$ magnitude. The effective temperature and surface gravity are taken as the calibrated values provided in APOGEE DR16. The $\alpha$-adjusted metallicity is calculated based on the calibrated metallicity ([M/H]) and $\alpha$-element abundance ([$\alpha$/M]) from APOGEE DR16 using the formula of \citet{Salaris1993}. The absolute $K$ magnitude is calculated using the 2MASS $K$ magnitude \citep{Skrutskie2006}, the distance from \citet{Bailer-Jones2018}, and the $K$-band extinction provided by APOGEE, AK\_TARG (or AK\_WISE if unavailable).

The prior on the SFH is flat in {$\rm \log(age)$} as all calculations are done in {$\rm \log(age)$}, and the PARSEC isochrones used are gridded in {$\rm \log(age)$} with step sizes of 0.05.  The MDF is assumed to be flat across the observational metallicity uncertainty, which is small compared to the typical spread of the disk MDF. We use the Chabrier lognormal IMF \citep{Chabrier2001} provided with the PARSEC isochrones. When analyzing a large sample of stars, a selection-function term is usually also included to account for the imposed limitations on surface gravity, color, and other parameters. However, the stars in the training sample in the current work are far from the edges of these selection cuts, and are minimally effected. 
A full description of the age determination method is described in \citet{Feuillet2016,Feuillet2018}. Note we do not perform the hierarchical modeling step employed by \citet{Feuillet2016} for the present work. 

\subsubsection{Validation}

The bottom row of Figure \ref{fig:apokasc} shows how well the [C/N]-[Fe/H] plane is covered by the training set of Feuillet ages for both the lower giant branch (2.6 $<$ $\log(g)$ $<$ 3.3, left panel) that overlaps with the APOKASC sample and the luminous giants ($\log(g)$ $<$ 2.0, right panel). The bottom-left panel qualitatively appears similar to the upper-left panel, indicating that ages are comparably mapped to the [C/N]-[Fe/H] space for both the APOKASC-2 seismic ages and the Feuillet ages for the lower giant branch stars. The right column emphasizes the lack of coverage using the APOKASC-2 ages as compared to the Feuillet ages, and also shows that the Feuillet ages map in a similar way to the [C/N]-[Fe/H] plane for both the lower and upper giant branches, confirming that there is age information in the C and N abundances for the luminous giants. Because the Feuillet ages rely on precise parallaxes of stars generally found in the MW disk, these two samples roughly sample the same volume of the Galaxy (inside of $\sim$ 3 kpc from the Sun).

Throughout this work, we only analyze stars with [Fe/H] $>$ $-$0.5. This is motivated by both theoretical studies of metallicity-dependent extra mixing along the giant branch (e.g., \citealt{Carbon1982,Charbonnel2010}) and recent empirical measurements of this extra mixing in the APOGEE data by \citet{Shetrone2019}. \citet{Shetrone2019} show that stars with $\log(g)$ $<$ 2.0 and [Fe/H] $<$ $-$0.5 can have extra mixing further affect the [C/N] abundance ratio by $\sim$ 0.15 dex at [Fe/H] = $-$0.7 and 0.58 dex at [Fe/H] = $-$1.4, making the [C/N] abundance ratio less sensitive to age. While we do not use the [C/N] abundance ratios explicitly to derive ages in this work, we know much of the age information comes from C and N features, likely making our ages susceptible to the same reduction in sensitivity.

An additional complication of deriving ages using carbon and nitrogen spectral features or [C/N] abundances is the difficulty of taking into account potential birth abundance variations in [C/N] abundances across the Galaxy. The APOKASC sample is confined to one region of the disk, and the Feuillet ages come from nearby stars (within $\sim$ 3 kpc from the Sun) by design, where the birth [C/N] abundance does not seem to vary by much (see e.g., \citealt{Martig2016a}). We discuss how a varying birth [C/N] in the bulge might affect our results in \S \ref{sec:disc_birth}.

\subsection{{\it The Cannon} with Feuillet Ages}
\label{sec:cannon}

Having verified that age information is contained in the APOGEE spectra for the luminous giants and that our training set spans a similar parameter space to the APOGEE bulge sample, we can use {\it The Cannon} \citep{Ness2015} to derive ages for these bulge stars. {\it The Cannon} is a data-driven technique for deriving stellar labels, where a model describing the flux at each pixel is created from a training set with well-known stellar labels. This model is then applied to a ``test'' set of spectra where the derived labels are returned. In this work, we fit a quadratic model to the spectra in our training set. While the [C/N] abundance could be used to map ages directly onto APOGEE stars via a multi-parameter fit, as was done in \citet{Martig2016a}, we opt to use the entire spectral range, as \citet{Ness2016b} showed that mass/age information is also encoded in the $^{12}$C/$^{13}$C ratio.

\subsubsection{Method}
\label{sec:training}

We build an input training sample from all stars with Feuillet ages (described in \S \ref{sec:train_set}) using the following cuts:

\begin{itemize}
    \item S/N $>$ 100 per pixel
    \item ~[M/H] $>$ $-$0.5
    \item No STAR\_BAD bit of ASPCAPFLAG flag set
    \item No STARFLAG bits set
    \item 0.5 $<$ $\log(g)$ $<$ 2.0
    \item {\it Gaia} DR2 parallax uncertainty $<$ 10\%
\end{itemize}

\begin{figure}[t]
\includegraphics[width=1.0\hsize,angle=0]{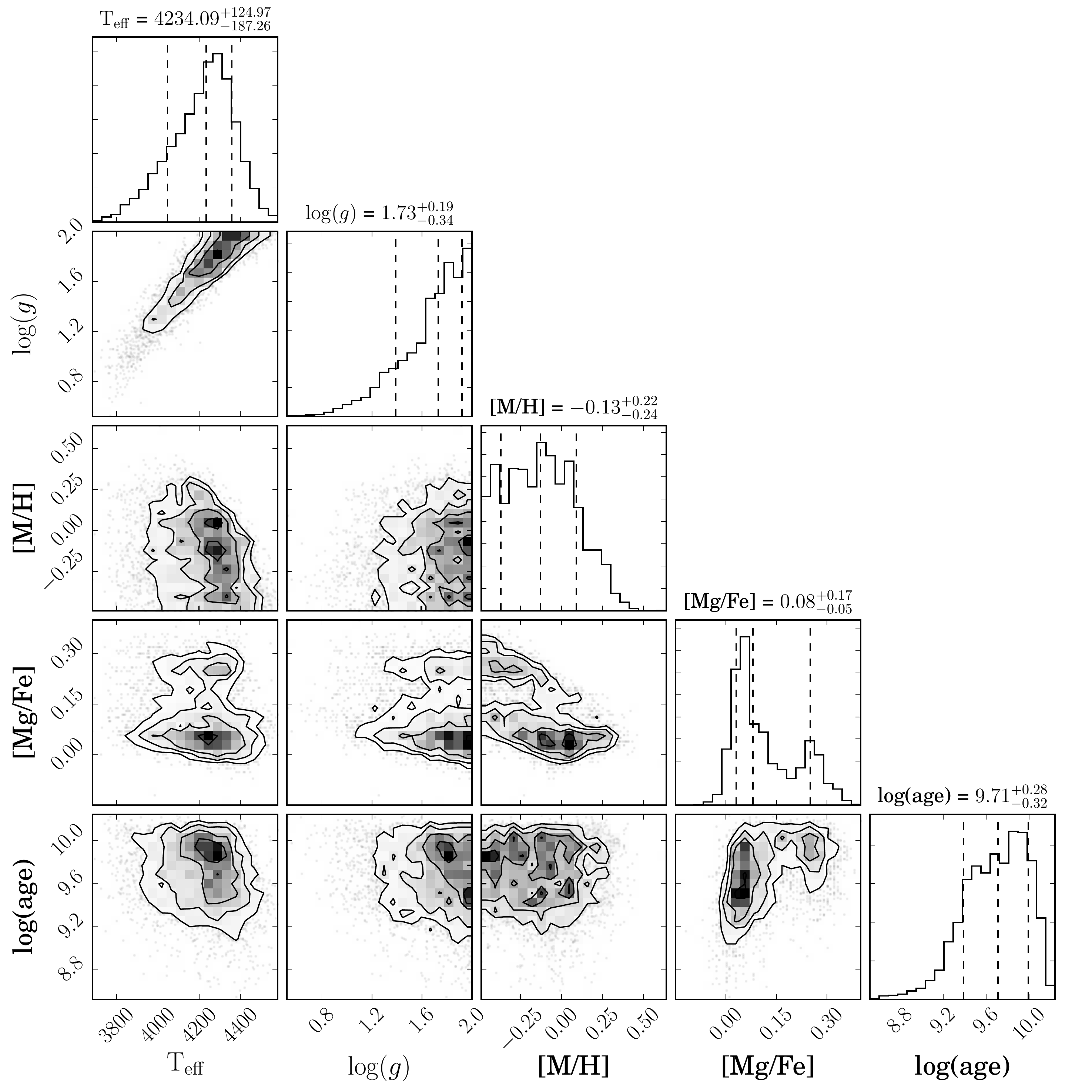}
\caption{Corner plot \citep{corner} of input labels for the training sample. The mean of each sample is indicated at the top of each row. Dashed lines in the 1D histograms mark the 0.16, 0.5, and 0.84 quantiles.
}
\label{fig:corner}
\end{figure}

These cuts result in a training sample of 3,711 stars that span the parameter space shown in Figure \ref{fig:corner}. We use {$\rm T_{eff}$}, $\log(g)$, [M/H], [Mg/Fe], and {$\rm \log(age)$} as the input labels. While the parameter space is reasonably well-covered, we note here that the ages will be extrapolated for stars with [M/H] $>$ +0.4 as the parameter space is barren this metal-rich. Therefore, ages for stars with [M/H] $>$ +0.4 should be used with caution. We run {\it The Cannon} using code obtained from Anna Ho\footnote{\url{https://github.com/annayqho/TheCannon}}, which also re-normalizes the APOGEE spectra. This additional normalization step changes the APOGEE-normalized spectra very little, but it does ensure that the spectra obtained from the Northern and Southern instruments are normalized in the same way.

\begin{figure*}[t]
\includegraphics[width=1.0\hsize,angle=0]{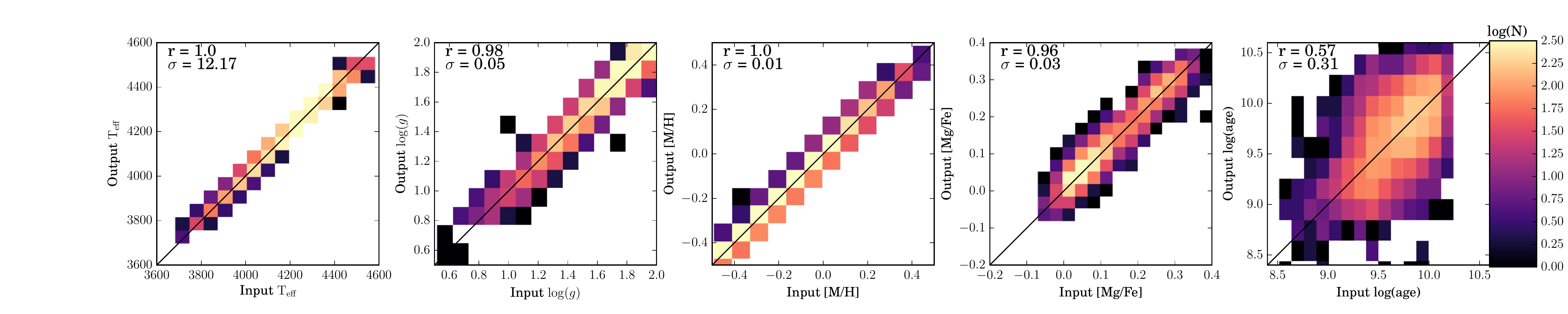}
\caption{Density plots showing the results of the 90-10 cross-validation test for each label. Pearson correlation coefficients and standard deviations are indicated in the upper-left of each panel.}
\label{fig:cross}
\end{figure*}

We first validate the output of the model by training {\it The Cannon} using 90\% of this training sample, then deriving labels for the remaining 10\%. We do this 10 times and analyze how the input labels compare to the output labels from this cross-validation test. The results are shown in Figure \ref{fig:cross}. Pearson correlation coefficients and standard deviations of the differences between the input and output labels are shown in the upper-left of each panel. 

We find that we are able to reproduce {$\rm T_{eff}$}, $\log(g)$, [M/H], and [Mg/Fe] to high precision, but the resultant ages are less precise, with a scatter around the 1--1 line of $\sim$0.3 dex, implying an uncertainty of $\sim$ 0.22 dex, which is slightly higher than other age studies using similar methods.
However, these are the first such ages where the training set sufficiently covers the bright end of the RGB, so that ages for our bulge stars are not extrapolations of the method. We further explore the age accuracy and precision in \S\ref{sec:age_results} and Appendix~\ref{sec:append_age_verify}.

Using this training set we run {\it The Cannon} on $\sim$46,000 luminous red giant stars in the APOGEE sample. These stars are selected to cover the same range of label space as the training set, specifically:

\begin{itemize}
    \item S/N $>$ 70 per pixel
    \item 0.5 $<$ $\log(g)$ $<$ 2.0
    \item $-$0.5 $<$ [Fe/H] $<$ $+$0.5
    \item No BAD bit of ASPCAPBAD flag set
\end{itemize}

When deriving labels for the test sample, we additionally remove any stars from the training sample that had reduced $\chi^{2}$ $>$ 2 in the cross-validation step, and retrain the model. Ages for all $\sim$ 46,000 stars along with the DR16 [Fe/H] and DR16 [Fe/H] uncertainties are provided in Table \ref{tab:ages}. The full table can be found in machine-readable format in the online journal. 

\begin{deluxetable*}{lcccccr}
\tablecaption{Ages produced in this work \label{tab:ages}}
\tablehead{
\colhead{APOGEE ID} & \colhead{APOGEE Field} & \colhead{Telescope} & \colhead{{$\rm \log(age)$}} & \colhead{$\sigma_{\rm log(age)}$} & \colhead{[Fe/H]} & \colhead{$\sigma_{\rm [Fe/H]}$}
}
\startdata
2M00000002+7417074 & 120+12 & apo25m & 9.36 & 0.30 & -0.17 & 0.01 \\
2M00000317+5821383 & 116-04 & apo25m & 9.68 & 0.28 & -0.28 & 0.01 \\
2M00000546+6152107 & 116+00 & apo25m & 9.02 & 0.24 & -0.27 & 0.01 \\
...\footnote{The full table can be found in machine-readable format on the online journal} & ... & ... & ... & ... & ... & ...\\
\enddata
\end{deluxetable*}

\subsubsection{Validation}
\label{sec:age_results}

The 90-10 test conducted in the previous section on the training sample suggests that we are reliably recovering age information from the APOGEE spectra. However, there are several external checks we can do to assess our age precision and accuracy further. Additional details can be found in \S \ref{sec:append_age_verify}, but we summarize the results here.

First we check for potential age dependence on telescope and instrument setup. Using a sample of 62 stars that were observed in both hemispheres for which we were able to derive ages, we find that, while most labels are identical, the age labels are offset such that the stars observed from LCO are 0.08 dex younger in {$\rm \log(age)$} as compared to the APO stars. Therefore, we apply a 0.08 dex offset to all LCO stars. We apply the offset in this direction because $\sim$ 75\% of our training set is comprised of stars observed from APO. This offset has already been applied to the ages reported in Table \ref{tab:ages}, and we note here that this offset does not affect our conclusions significantly.

The results of the cross-validation test (Figure \ref{fig:cross}) suggest a precision in {$\rm \log(age)$} of $\sim$ 0.22 dex. After exploring the potential dependence of this precision on S/N, [M/H], [Mg/Fe], and $\log(g)$, we find that the precision only depends on $\log(g)$, with the lower-$\log(g)$ stars being less-precise. In \S \ref{sec:append_uncert} we describe how we derive a quadratic fit to {$\rm \log(age)$} random uncertainties as a function of $\log(g)$. These uncertainty values are reported in Table \ref{tab:ages}, and range from 0.3 dex precision for the lowest $\log(g)$ values ($\log(g)$ $\simeq$ 0.5-0.8) to 0.2  dex precision for the highest $\log(g)$ values ($\log(g)$ = 2.0).

For a first check of the accuracy of our age results we cross match our age catalog to the open cluster catalog of \citet{Cantat-Gaudin2018}. We find 3 clusters for which we have derived ages for $>$ 3 members: NGC 6791, NGC 6819, and NGC 2204. We find that the ages we derive using the median of the ages of cluster members agree to $\sim$ 0.1 dex or better for the youngest two clusters, but we find a median age that is 0.2 dex higher than the age reported in the \citet{Kharchenko2013} catalog for NGC 6791. These results are explained in more detail in \ref{sec:append_cluster}.

We can also check the accuracy of our results by comparing stars that have ages both from {\it The Cannon} and APOKASC, limited to 2.0 $<$ $\log(g)$ $<$ 1.5. We find that the ages agree reasonably well where the APOKASC mass uncertainties are low, but when the APOKASC mass uncertainties become larger than $\sim$ 10\%, the APOKASC ages are systematically larger than the ages we derive using {\it The Cannon}. This is discussed in more detail in \S \ref{sec:append_apokasc}, but given that the offset is only for stars with large APOKASC uncertainties, and on the upper end of the $\log(g)$ range considered, we do not apply any offset to our data. 

For another check on the accuracy of our ages, we reproduce age-abundance maps of the MW that have been previously studied in works such as \citet{Ness2016b} and \citet{Martig2016b}. These maps are shown and described in further detail in \S \ref{sec:append_disk}. Although we study more luminous giants with ages derived from a different training set than all previous work, we find very similar qualitative trends first found and described by \citet{Ness2016b} and \citet{Martig2016b}. We also sub-select our sample to match the spatial distribution of the CoRoGEE sample studied by \citet{Anders2017b} and measure near-identical metallicity gradients for stars younger than 2 Gyr and stars older than 10 Gyr. 

Finally, in \ref{sec:append_dwarfs}, we show the ages we are able to get for the most metal-rich stars in the Large Magellanic Cloud (LMC) and Sagittarius Dwarf (Sgr) galaxies. We find that the most metal-rich LMC stars have a median age of $\sim$ 0.8 Gyr, consistent with the expected ages of these stars from star formation history studies. Similarly, we find the Sgr stars to have age $\sim$ 6 Gyr, again consistent with what is expected from the literature.

\section{Results}
\label{sec:results}

We now present our age results, first showing mean age maps for much of the entire APOGEE sample, then focusing on the bulge sample defined in \S \ref{sec:obs}.

\subsection{Mean-Age Maps}

\begin{figure*}[t]
\includegraphics[width=1.0\hsize,angle=0]{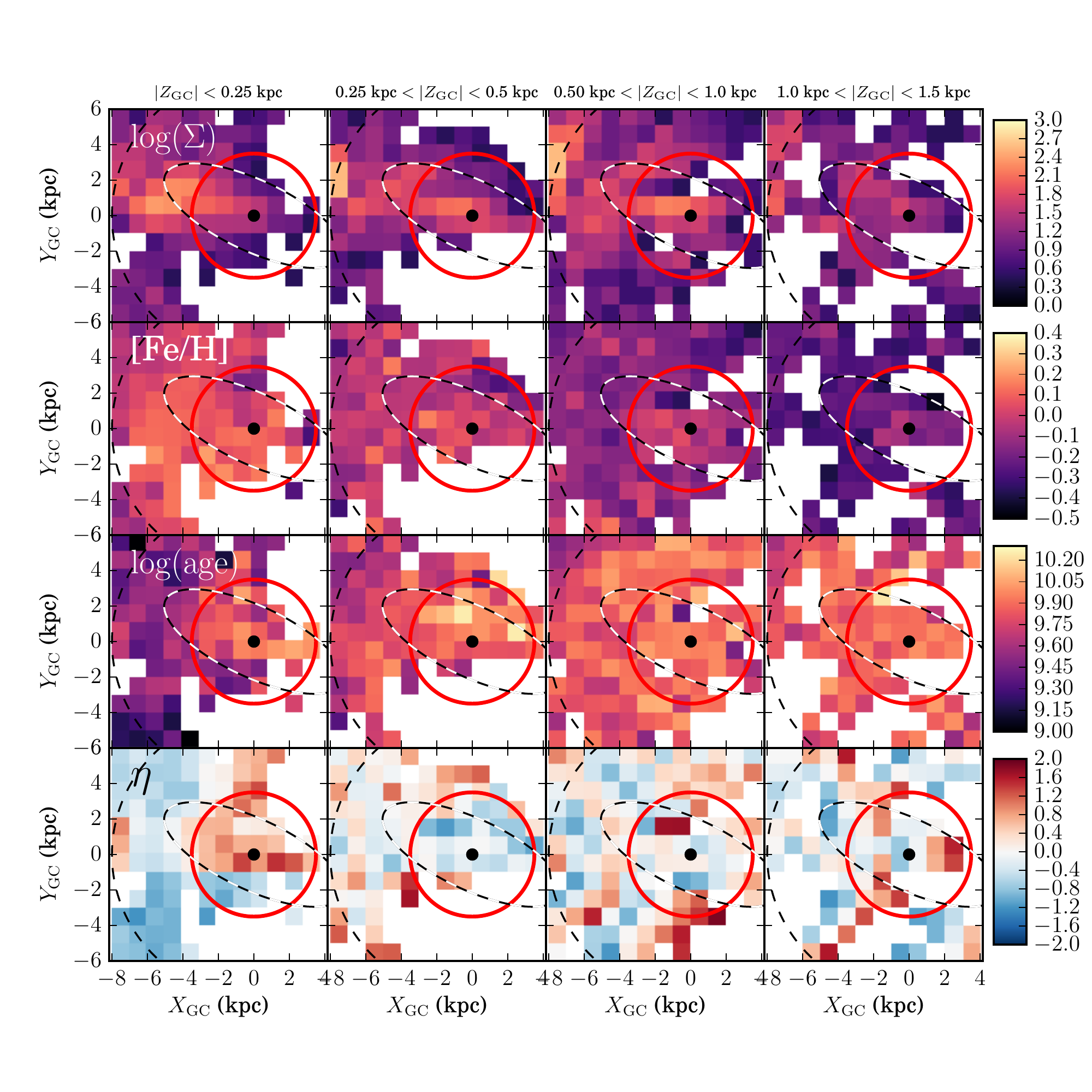}
\caption{Maps of the MW for different $|Z_{\rm GC}|$ bins, separated by columns, and colored by stellar density (top row), [Fe/H] (second row), {$\rm \log(age)$} (third row), $\eta$, described in detail in the text. The red solid circle denotes the $R_{\rm cy} =$ 3.5 kpc bulge selection. The black dashed circle marks the solar circle. The black and white dashed ellipse marks the bar as described in \citet{Bovy2019}.}
\label{fig:maps}
\end{figure*}

First we introduce some qualitative spatial age trends of the bulge. In Figure \ref{fig:maps} we show maps of the $X_{\rm GC}-Y_{\rm GC}$ plane of the MW for different $|Z_{\rm GC}|$ bins colored by number density of stars ($\Sigma$, top row), [Fe/H] (second row), {$\rm \log(age)$} (third row), and $\eta$ (fourth row), which is the number of standard deviations each bin $i$ is away from the mean-$|Z_{\rm GC}|$ of that entire $|Z_{\rm GC}|$-range that divides the columns. This serves as a metric to assess potential spatial biases induced in the binning.

\begin{equation}
\label{eqn:eta}
\eta=\frac{<Z_{\rm GC}>_{i}-<Z_{\rm GC}>_{Tot}}{\sigma_{|Z_{\rm GC}|}}
\end{equation}

The top row of Figure \ref{fig:maps} shows the stellar density at each position in the Milky Way, to demonstrate how APOGEE samples the bulge at these metallicities and $\log(g)$ values (also see \citealt{Bovy2019} and \citealt{Queiroz2020}). As expected, the near side of the bulge ($X_{\rm GC}$ $<$ 0 kpc) is covered at a much higher stellar density than the far side ($X_{\rm GC}$ $>$ 0 kpc). The far side also has relatively incomplete $X_{\rm GC}$ and $Y_{\rm GC}$ coverage, especially for the two bins with $|Z_{\rm GC}|$ $<$ 0.5 kpc. This is not surprising given the extinction in the midplane, where $A_V$ frequently exceeds 25 mag (e.g., \citealt{Schultheis1999}). Therefore, for all future analysis discussed in this work, we restrict our sample to be on the near side of the bulge, or $X_{\rm GC}$ $<$ 0 kpc.

The second row of Figure \ref{fig:maps} shows how the mean [Fe/H] of our sample, which is limited to [Fe/H] $>$ $-$0.5, changes with position in the Galaxy. As has been found before for stars at $R_{\rm cy}$ $<$ 8 kpc, the mean [Fe/H] decreases with increasing $|Z_{\rm GC}|$ (e.g., \citealt{Hayden2015}). Similar to \citet{Leung2019}, we find that the metallicity of the MW appears to peak around $R_{\rm cy}$ $\sim$ 4-5 kpc, and maybe even decreases in the inner-most region. We also see that the region of the Galaxy with 0 kpc $<$ $Y_{\rm GC}$ $<$ 3 kpc and $-$3 kpc $<$ $X_{\rm GC}$ $<$ 0 kpc appears to be more metal poor than other regions of the bulge. This low-metallicity feature was also seen by \citet{Leung2019} and \citealt{Queiroz2020:inprep}. \citet{Bovy2019} interpreted this as a signature of the bar, which also was shown to be distinct in age and kinematics. However, \citet{Wegg2019} actually find the bar to be more metal rich than the disk. This is potentially a result of incomplete $|Z_{\rm GC}|$ sampling, described more below.

The third row of Figure \ref{fig:maps} shows how the mean {$\rm \log(age)$} changes with position in the Galaxy. Outside of 3 kpc stars appear to exhibit a steeper vertical age gradient as compared to stars inside of 3 kpc, which appear to all be old. This is in qualitative agreement with studies finding the mean age of bulge stars with [Fe/H] $>$ $-$0.5 to be old. This results in a radial age gradient where the ages of stars in the plane ($|Z_{\rm GC}| <$ 0.25 kpc) go from {$\rm \log(age)$} $\sim$ 9.8-9.9 at $R_{\rm cy} < $ 3 kpc to {$\rm \log(age)$} $\sim$ 9.3 at $R_{\rm cy} = $ 5 kpc. Out of the plane, there appears to be no radial age gradient from 0 kpc $< R_{\rm cy} <$ 8 kpc. In the planar bin, $|Z_{\rm GC}|$ $<$ 0.25 kpc, we also find signs of the off-bar side of the bulge ($l$ $<$ $0^{\circ}$, $Y_{\rm GC}$ $<$ 0 kpc) being younger, on average, than stars on the on-bar side of the bulge ($l$ $>$ $0^{\circ}$, $Y_{\rm GC}$ $<$ 0 kpc), as was found by \citet{Bovy2019}. We further explore and quantify these differences in \S \ref{sec:results_onoff}.

However, the interpretation of such gradients, as well as on-bar vs. off-bar comparisons, are complicated by potential selection biases. We show in the bottom row of Figure \ref{fig:maps} the same maps but colored by $\eta$, defined as the number of $\sigma(Z_{\rm GC})$ values away from the mean of the entire $|Z_{\rm GC}|$ range (Equation~\ref{eqn:eta}). $X_{\rm GC}-Y_{\rm GC}$ bins that have little color have stars with a mean $|Z_{\rm GC}|$ close to the mean, whereas bins that are dark red or dark blue have stars that are up to 2$\sigma$ away from the mean. In the case for the $|Z_{\rm GC}|$ $<$ 0.25 kpc bin, we show that the bulge is probed at a larger mean $|Z_{\rm GC}|$ than the disk region ($R_{\rm cy} >$ 4.0 kpc), and different sides of the bulge are probed at different mean $|Z_{\rm GC}|$ heights. Therefore, any intrinsic vertical age/metallicity gradients that exist in the inner Galaxy will potentially cause one to measure different mean ages/metallicities if the $|Z_{\rm GC}|$-sampling is not taken into account. We discuss how this affects spatial variations in the age distribution of the bulge in \S \ref{sec:results_onoff}.

\subsection{Age-Metallicity Relation of the Bulge}

For the remainder of this section, we will focus on the bulge stars only, which we define as:

\begin{itemize}
    \item $R_{\rm cy} <$ 3.5 kpc
    \item $|Z_{\rm GC}|$ $<$ 1.5 kpc
    \item $X_{\rm GC}$ $<$ 0 kpc (near side of MW only)
\end{itemize}

\begin{figure*}[t]
\includegraphics[width=1.0\hsize,angle=0]{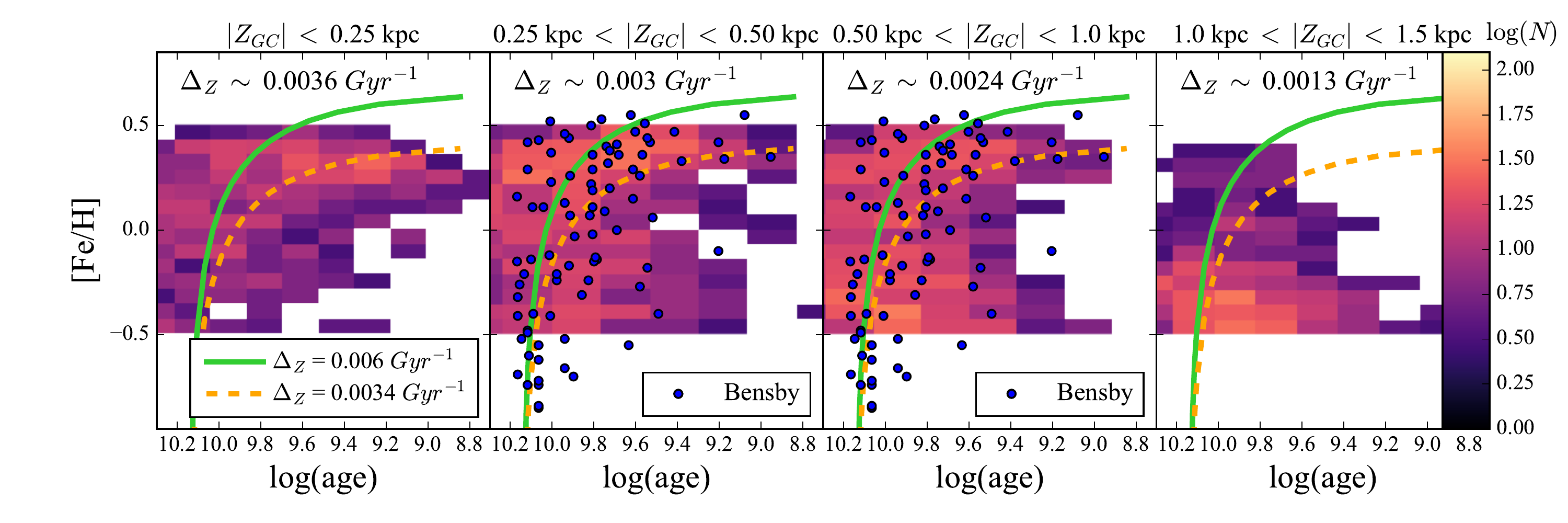}
\caption{Age-[Fe/H] relations for four different $|Z_{\rm GC}|$ bins. Stars from \citet{Bensby2017} are over-plotted only in the central two panels where they are likely to overlap spatially with the APOGEE sample. The two lines in each panel show the linear metal enrichment, $\Delta_{Z}$, for the fit to the \citet{Bernard2018} sample (green solid line), and the \citet{Haywood2016} model (orange dashed line). The upper region of each panel shows the $\Delta_{Z}$ we calculate for the APOGEE data, from 13.7 to 7 Gyr ago.}
\label{fig:age_feh_relation}
\end{figure*}

The reason that we restrict our sample to the near side of the bulge is because of the ``patchy'' nature of the APOGEE bulge coverage, as shown in Figure \ref{fig:maps}. Figure \ref{fig:age_feh_relation} shows the age-metallicity relation for these stars in the same $|Z_{\rm GC}|$ bins shown in Figure \ref{fig:maps}. In the upper panel we calculate the linear metal enrichment over time, $\Delta_{Z}$ = $dZ/dt$, where $Z$ here is metallicity mass fraction, \textbf{not} Galactic height, $Z_{\rm GC}$. To calculate this value, we find the mean metallicity of stars with 9.75 $<$ {$\rm \log(age)$} $<$ 9.95 in each $Z_{\rm GC}$ bin, and assume the stars enriched from metallicity $Z$ = 0 at $t$ = 13.7 Gyr to the $Z$ observed at this age range, which is $\sim$ 7 Gyr ago. We choose this age to compare to other works that typically calculate this enrichment over this time period. This age limit is usually imposed because after 7 Gyr, the stars enrich very slowly over time (e.g., \citealt{Bernard2018}), or are not even found in other samples.

We find that the metallicity evolution of the bulge was quicker for the stars closer to the plane than for stars out of the plane. We also find that our stars with $Z_{\rm GC}$ $<$ 0.5 kpc have $\Delta_{Z}$ values consistent with the model put forth by \citet{Haywood2016}, but shallower than the fit for the same age range in \citet{Bernard2018}. 

We also over-plot the \citet{Bensby2017} stars in the middle two bins, corresponding to the most likely spatial overlap given the latitude of the \citet{Bensby2017} stars. The ages of these stars are not inconsistent with what we find in our sample. We also see that the super-solar metallicity stars extend down to the lower ages that Bensby et al. observed. As described more in \S \ref{sec:young_stars}, we find that the mean age of the stars with $+$0.2 $<$ [Fe/H] $<$ $+$0.4 is $\sim$ 9.5-9.6 in {$\rm \log(age)$}, or 3-4 Gyr old. This is only true for the spatial bin closest to the plane. 

We also find signs that the ages of the most metal-poor stars ($-$0.5 $<$ [Fe/H] $<$ $-$0.3) in our sample become slightly younger, on average, at larger distances from the plane. \citet{Bernard2018} find a reasonable spread in age at these metallicities, suggesting that finding stars at $-$0.5 $<$ [Fe/H] $<$ $-$0.3 that are as young as 7-8 Gyr old is not unusual. The fact that these stars are younger farther from the plane also fits with overall slower metallicity evolution farther from the plane, and represent the metallicity of the last stars formed at these $|Z_{\rm GC}|$-heights.

\subsection{Young Stars in the Bulge}
\label{sec:young_stars}

The age-metallicity relations suggest that the bulge is not uniformly old, and we find stars as young as $\sim$ 1-3 Gyr at [Fe/H] $>$ 0.1 and $|Z_{GC}| <$ 0.25 kpc. To assess the significance of these stars, given our age uncertainties, we divide our sample into mono-abundance bins (e.g., \citealt{Bovy2016}), and analyze the age distribution of each mono-abundance bin. Specifically, we are interested in the median age of each mono-abundance bin, and whether each mono-abundance bin can be described by a single age.

\begin{figure*}[t]
\includegraphics[width=1.0\hsize,angle=0]{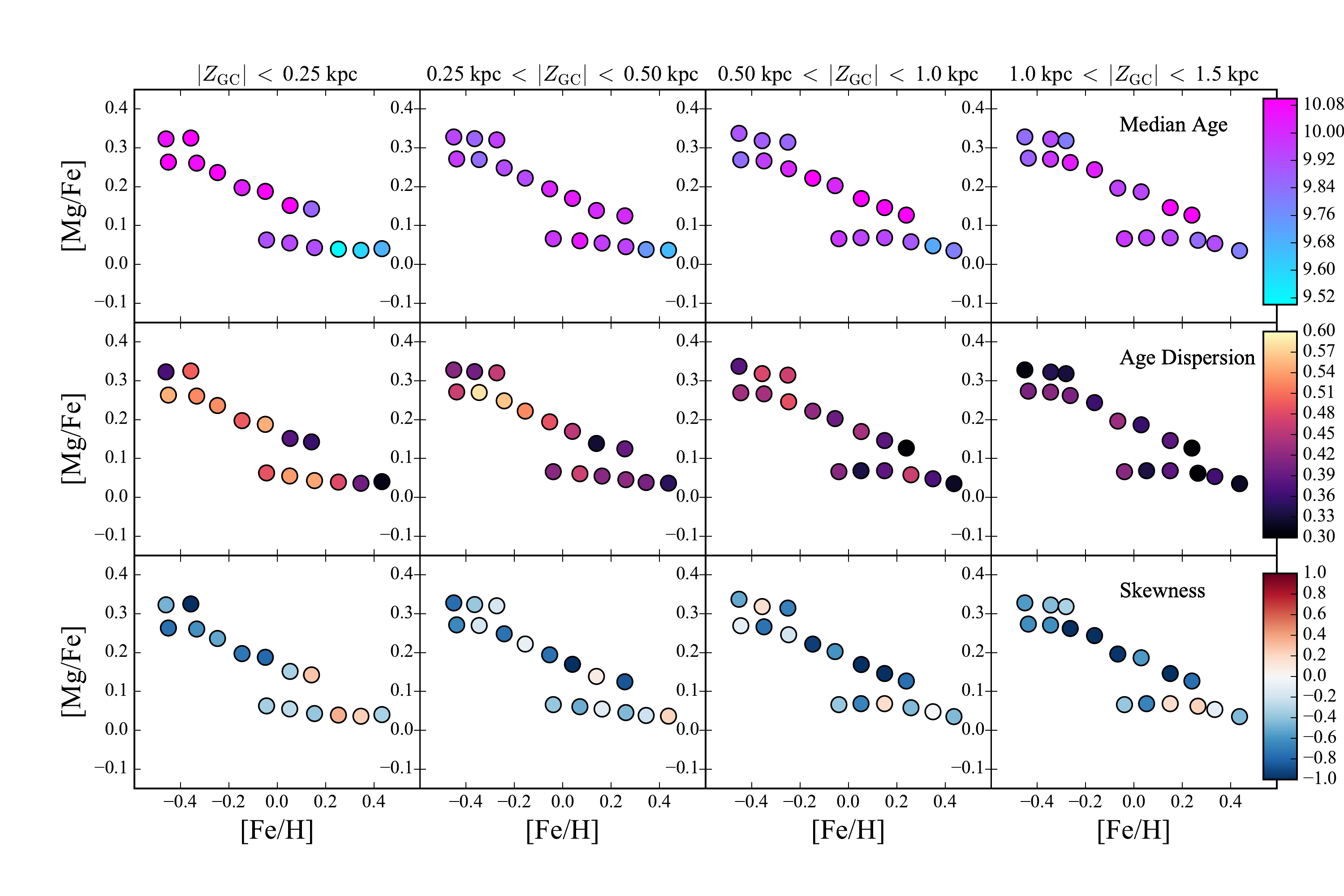}
\caption{Mono-abundance bins across $|Z_{\rm GC}|$ bins colored by median age (top row), age dispersion (middle row), and skewness (bottom row). Only mono-abundance bins containing more than 20 stars are plotted. Stars are divided in 0.1 dex bins of [Fe/H] and 0.2 dex bins of [Mg/Fe], and the medians of the stars in each bin are plotted as points.}
\label{fig:mean_age}
\end{figure*}

Figure \ref{fig:mean_age} shows the median age (top row), standard deviation in age (second row), and skewness in age (third row) for each mono-abundance bin for the four $|Z_{GC}|$ bins. Stars are divided into 0.1 dex bins of [Fe/H], from $-$0.5 $<$ [Fe/H] $<$ $+$0.5, and 0.2 dex bins of [Mg/Fe] from $-$0.1 $<$ [Mg/Fe] $<$ $+$0.5. Points are plotted according to the median [Fe/H] and [Mg/Fe] of each bin, so therefore do not necessarily lie on a grid point. Points are only plotted if the bin contains more than 20 stars.

As shown in the top row of Figure \ref{fig:mean_age}, we find the bins containing stars with [Fe/H] $>$ $+$0.2 and [Mg/Fe] $<$ $+$0.1 are generally the youngest in median age, with the age of these populations slightly increasing from in the plane ({$\rm \log(age)$} $\sim$ 9.5) to out of the plane ({$\rm \log(age)$} $\sim$ 9.8). While stars outside these abundances are generally all older with {$\rm \log(age)$} $>$ 9.9, there is potentially a slight age gradient, such that stars with [Fe/H] $<$ $-$0.2 are $\sim$ 0.1 dex younger out of the plane than stars in the plane.

We also show the standard deviation of {$\rm \log(age)$} for each mono-abundance population in the middle row of Figure \ref{fig:mean_age}. Given our age uncertainties of 0.22-0.30 dex, points that are colored black are consistent with being comprised of stars of the same abundance and age. Points that are lighter in color likely have a real intrinsic age spread. We find nearly no age spread for mono-abundance bins at large distances from the plane, but substantial age dispersion for mono-abundance bins in the plane, particularly at [Fe/H] $<$ 0.0 and super-solar [Fe/H] stars with [Mg/Fe] $<$ $+$0.1. In the plane, the populations with [Fe/H] $>$ $+$0.2 and [Mg/Fe] $<$ $+$0.1, which are the population with the youngest median age, actually exhibit smaller age dispersion than the more metal-poor stars also with [Mg/Fe] $<$ $+$0.1. 

Finally, the third row of Figure \ref{fig:mean_age} shows the same mono-abundance bins, but now colored by the skewness of the age distribution. Dark blue points correspond to abundance bins that contain a skew towards younger ages and dark red points correspond to abundance bins that contain a skew towards older ages. The two youngest bins in the plane actually exhibit a slight positive skew, indicating that these abundance bins still contain old stars. The stars at large distances from the plane with [Fe/H] $<$ $-$0.1 exhibit large negative skew, suggesting that these more metal-poor stars are not uniformly old, and contain a smaller fraction of younger stars, as also found by \citet{Bernard2018}.

\begin{figure*}[t]
\includegraphics[width=1.0\hsize,angle=0]{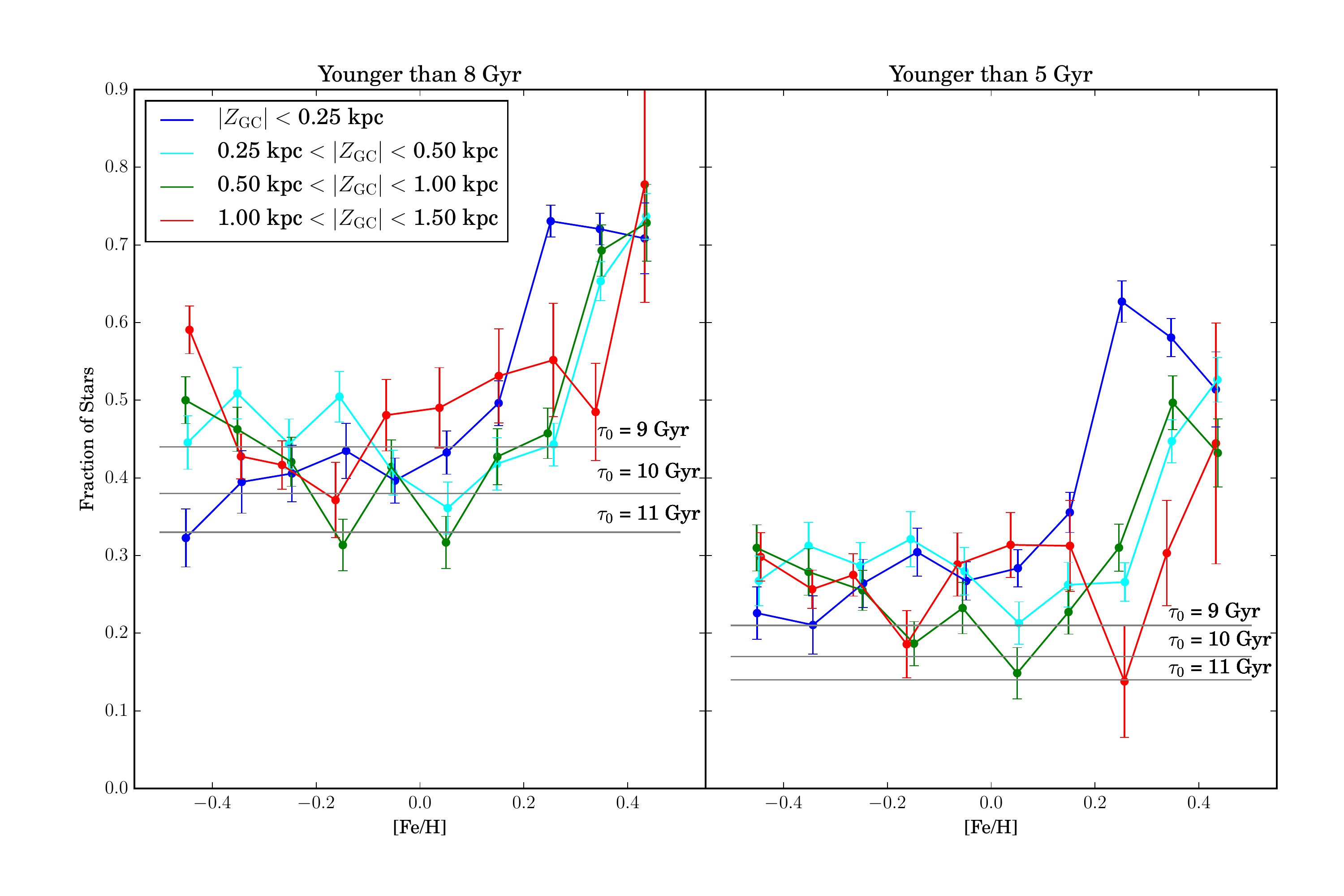}
\caption{Fraction of stars in the bulge sample younger than 8 Gyr (left panel) and 5 Gyr (right panel) as a function of [Fe/H] for four different ranges of  $|Z_{\rm GC}|$-heights (colored lines). The gray horizontal lines in each panel indicate the expected fraction, given our age uncertainties, if the stars were born all at 9, 10, and 11 Gyr ago ($\tau_{0}$). }
\label{fig:frac_young}
\end{figure*}

In Figure \ref{fig:frac_young} we summarize the $|Z_{\rm GC}|$-heights and [Fe/H], where excess young stars can be found in the bulge. We plot the fraction of stars younger than 8 Gyr and 5 Gyr, as was done in \citet{Bernard2018}, but do this for a range of $|Z_{\rm GC}|$-heights. All $|Z_{\rm GC}|$ bins exhibit an increase in the fraction of younger stars with increasing [Fe/H], but this increase starts at lower [Fe/H] for stars closer to the plane. This is true for both 8 Gyr and 5 Gyr. The gray lines in each panel of Figure \ref{fig:frac_young} show the expected fraction, given our age uncertainties, if the stars in each [Fe/H] bin were formed at a single age of 9, 10, and 11 Gyr, with a 0.1 dex spread in $\log({\rm age})$. Also, the left panel of Figure \ref{fig:frac_young} shows that stars with $|Z_{\rm GC}|$ $>$ 0.5 kpc and [Fe/H] $<$ $-$0.3 have a non-neglible fraction of stars with age $<$ 8 Gyr, potentially a consequence of the overall slower chemical evolution farther from the plane.

So while many of the stars, especially at solar [Fe/H] and below, appear to be consistent with being born from one event some 9-10 Gyr ago, we measure a significant fraction of younger stars for the more metal-rich stars, where ``young'' means age $<$ 5 Gyr. However, we note that the youngest stars in our full MW age sample at these metallicities are actually found outside of the bulge ($R_{\rm cy}$ $\sim$ 5 kpc, see \S \ref{sec:append_disk}). Therefore, while we see little evidence for very recent star formation in the bulge, we do find strong evidence that the bulge formed stars as recently as 2-5 Gyr ago, and that this star formation took place at $+$0.2 $<$ [Fe/H] $<$ $+$0.4 and in the plane.

\begin{figure}[t]
\includegraphics[width=0.95\hsize,angle=0]{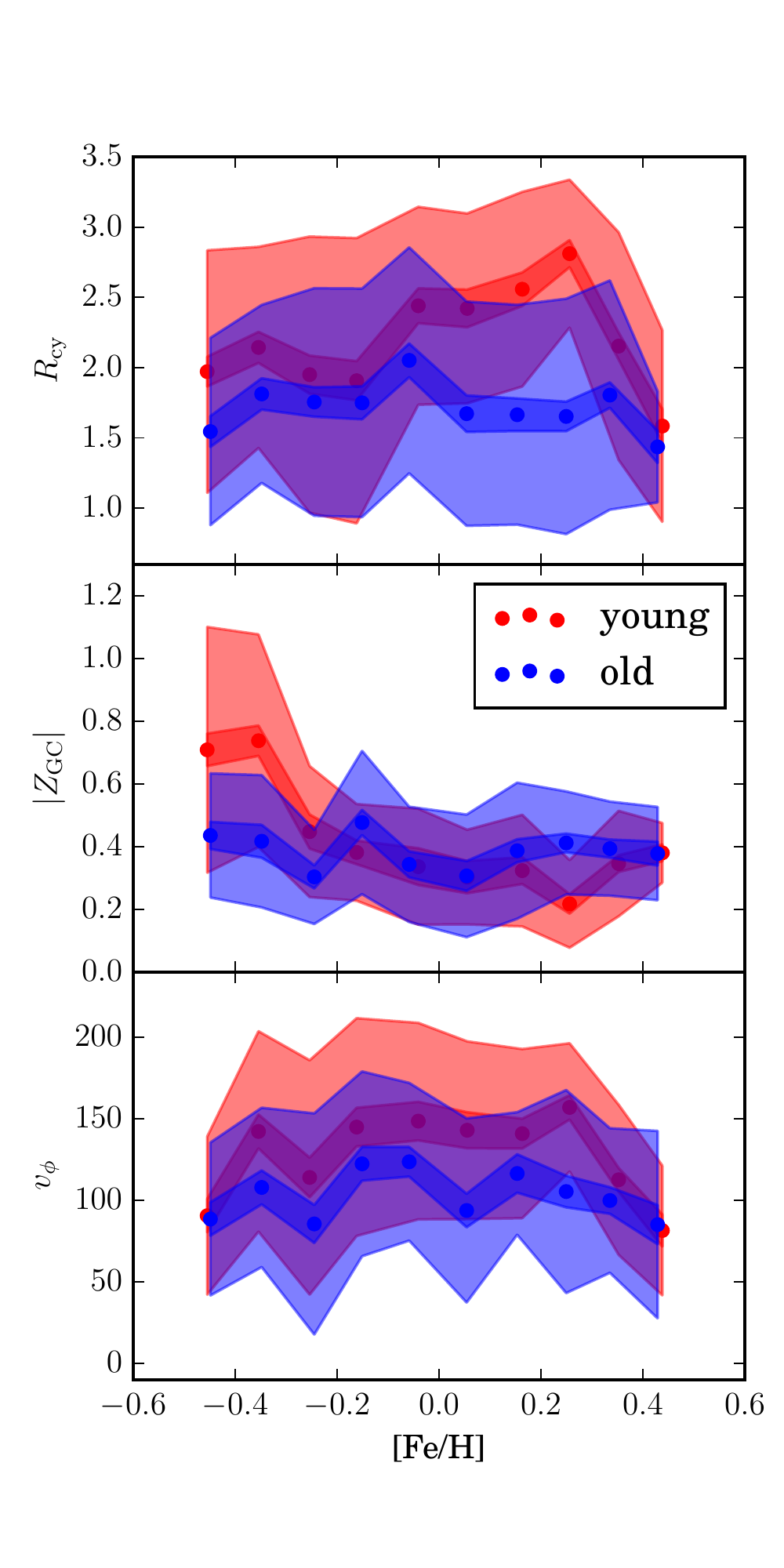}
\caption{Median $R_{\rm cy}$, $|Z_{\rm GC}|$, and $v_{\phi}$, as a function of [Fe/H] for 0.1 dex bins of [Fe/H]. ``Young'' and ``Old'' stars, as described in the text, are plotted as red and blue, respectively. The lighter shaded regions show the range of median absolute deviation, and  the darker shaded regions show the range of standard errors on the median.}
\label{fig:young}
\end{figure}

To further analyze the spatial and kinematical properties of the younger stars we find in the bulge, we divide the bulge sample into 0.1 dex bins of [Fe/H], and define an ``old'' and ``young'' sample for each bin, where ``old'' stars are stars greater than one standard deviation away from the mean age of each bin, and ``young'' stars are stars less than one standard deviation away from the mean of each bin. The median $R_{\rm cy}$, $|Z_{\rm GC}|$, and $v_{\phi}$ of these stars are plotted as a function of median [Fe/H] in Figure \ref{fig:young}. 

The first row of Figure \ref{fig:young} shows that the young stars with $+$0.1 $<$ [Fe/H] $<$ $+$0.3 tend to be found at larger Galactocentric radii than the old stars at the same metallicity, and at slightly lower $|Z_{GC}|$ (second row of Figure \ref{fig:young}). The young stars at these metallicities also tend to have larger rotational velocities than the older stars (third row of Figure \ref{fig:young}), suggesting that the young stars are generally still in a disk as compared to the old stars at those metallicities. Again, we find that the younger metal-poor stars are typically found at higher $|Z_{\rm GC}|$ than the older stars.

To summarize, while we do indeed find that many stars in the bulge at the metallicities studied here are 9-10 Gyr old, there is a statistically significant number of metal-rich stars younger than 5 Gyr that tend to be at 2 $<$ $R_{cy}$ $<$ 3 kpc, $|Z_{\rm GC}| <$ 0.25 kpc , and exhibit kinematics consistent with rotating around the MW center. Therefore, while the bulk of the metal-rich bulge formed in one event some 9-10 Gyr ago, star formation continued in a disk at super-solar metallicities until $\sim$ 2 Gyr ago.

\subsection{On and Off the Bar}
\label{sec:results_onoff}

Because APOGEE-2 has observed stars across the entire bulge, we are able to compare stars on the on-bar side of bulge to those on the off-bar side. To select these stars, we follow the lead of \citet{Bovy2019} and put stars in the on-bar sample if they fit inside an ellipsoidal structure oriented at 25$^{\circ}$ from the Sun, with a half-width of 2 kpc. The off-bar sample consists of the stars that fall outside of this region. This ellipse used to denote on- and off-bar stars is shown on the maps of Figure \ref{fig:maps}. To account for potential radial variation, we also limit both samples to be 2.0 $<$ $R_{\rm cy}$ $<$ 3.5 kpc, and as before, only consider stars at $X_{\rm GC}$ $<$ 0 kpc.

\begin{figure}[t]
\begin{center}
\includegraphics[width=0.95\hsize,angle=0]{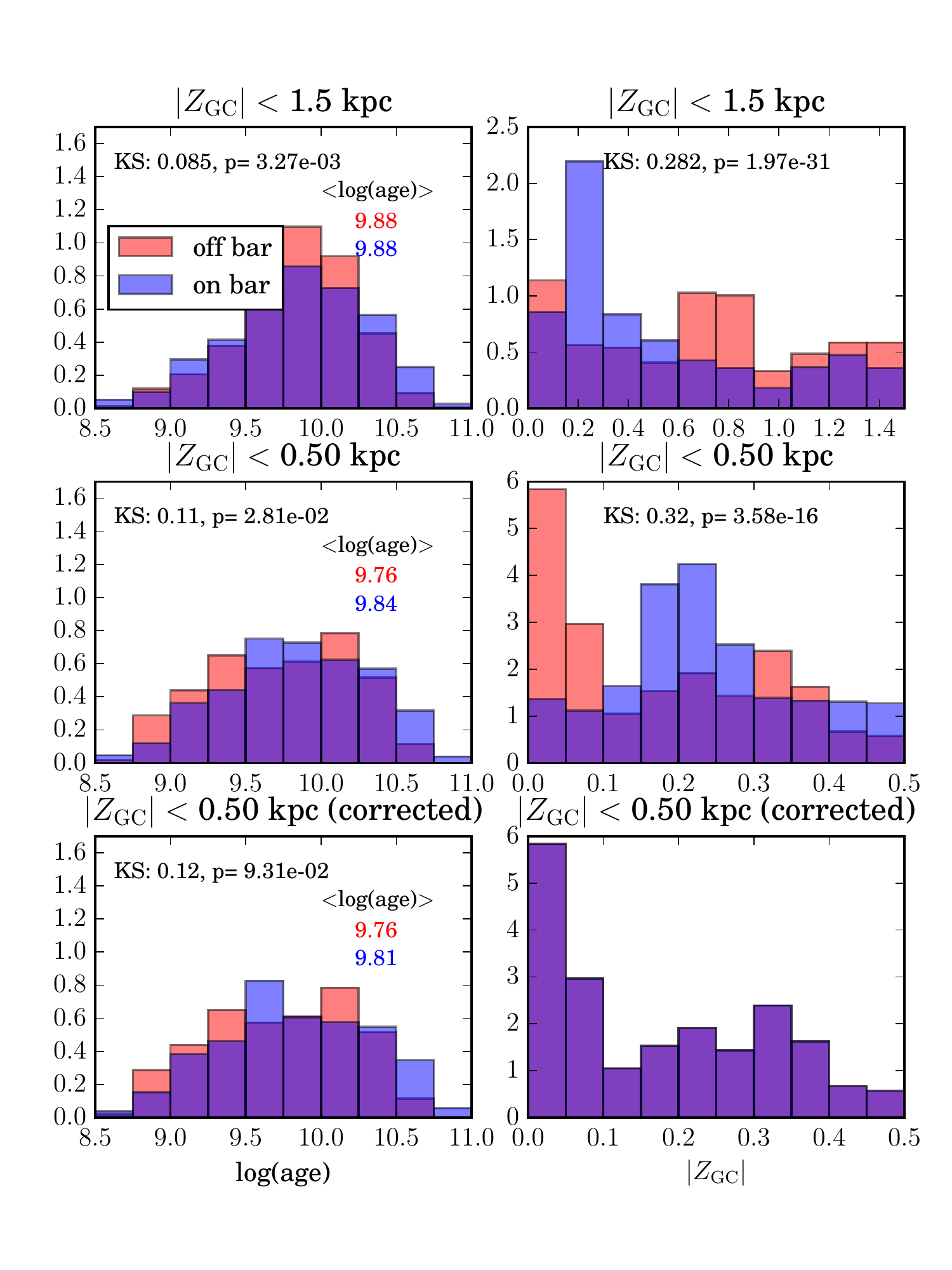}
\caption{Age distribution and $|Z_{\rm GC}|$ distributions separately for the on-bar (blue) and off-bar (red) samples in bins of $|Z_{\rm GC}|$. Results of KS tests are indicated in the upper-left of each panel. Median {$\rm \log(age)$} values are included in the left-column panels.}
\label{fig:onoff}
\end{center}
\end{figure}

The mean maps shown in Figure \ref{fig:maps} suggest that the off-bar side is more metal-rich and slightly younger than the on-bar side, but only at $|Z_{\rm GC}|$ $<$ 0.50 kpc. We measure these differences and quantify their significance in Figure \ref{fig:onoff}. The top row shows the age distribution of the stars in the off-bar (red) and on-bar (blue) samples defined above. While the median ages of the two samples are nearly identical, the results of a KS test suggest a small, but potentially significant difference in the age distributions. The differences become more pronounced when we limit the sample to $|Z_{\rm GC}|$ $<$ 0.50 kpc (middle row of Figure \ref{fig:onoff}). However, the right panel highlights that there is also a difference in the $|Z_{\rm GC}|$ distributions probed, leaving open the possibility of sampling biases.

We correct for this by sub-sampling the on-bar sample, which contains $\sim$ 10 times more stars than the off-bar sample, such that it matches the same $|Z_{\rm GC}|$ distribution as the on-bar sample. The resultant age distributions are shown in the bottom row of Figure \ref{fig:onoff}. With this correction, we find that the KS test cannot rule out the possibility that these two distributions are drawn from the same parent distribution. More data from the Southern Hemisphere telescope will help to further quantify the similarities (or differences) of these on- and off- bar age distributions distributions, as well as more careful separation into on- and off-bar groups using orbital information (see \citealt{Queiroz2020:inprep}).

The above results also suggest that the apparent old age of the bar shown in Figure \ref{fig:maps} is largely or entirely a sampling effect, with the stars in the inner 0.1 kpc of the plane exhibiting younger ages, on average, than the rest of the stars. Additionally, because the mean age of the on-bar sample decreases when we sub-sample to match the $|Z_{\rm GC}|$ distribution of the off-bar sample, then there is likely at least a slight vertical age gradient. Quantifying this vertical age gradient is beyond the scope of this work, as careful evaluation of selection effects would need to be taken into account.

\section{Discussion}
\label{sec:disc}

\subsection{Spatial Age Trends of the Bulge}

We find that $\sim$ 50\% of our bulge sample, comprised of stars with $-$0.5 $<$ [Fe/H] $<$ $+$0.5, are older than 8 Gyr. As shown in Figure \ref{fig:frac_young}, nearly all stars with sub-solar metallicity have ages consistent with being born from one star-formation event some 9-10 Gyr ago. This agrees with many lines of evidence that point to an old bulge, including the chemical track of bulge stars in [$\alpha$/Fe]-[Fe/H] space (e.g., \citealt{Cunha&Smith2006,Fulbright2007,Johnson2011,Hill2011,McWilliam2016,Bensby2017,Zasowski2019,Bovy2019}), where most or all of the sub-solar metallicity stars are shown to be enhanced in their $\alpha$-elements. Recent work suggests that the chemistry of the bulge stars are identical to the chemistry of the thick disk stars found in the Solar Neighborhood (e.g., \citealt{Rojas-Arriagada2014,Bovy2019}). These stars are found to be at similar ages to what we find for our bulge stars at sub-solar metallicities, serving as further evidence for a coeval formation scenario.

We do find some spatial variations in the age distributions of the bulge. First, we find a slight vertical age gradient from 0 $<$ $|Z_{\rm GC}|$ $<$ 0.5 kpc, where the stars closer to the plane are $\sim$ 0.2 dex younger than stars farther from the plane. A much steeper vertical gradient is observed outside of the bulge at $R_{\rm cy}$ $\sim$ 5 kpc, which is driven by the inclusion of the youngest stars (0.5-2 Gyr) in the plane. These youngest stars are largely not present in the bulge sample. We find that the bulge enriched at a rate of $dZ/dt \sim$ 0.0034 ${\rm Gyr^{-1}}$, which is very similar to the rate predicted by \citet{Haywood2018}, but we find that this rate decreases with height above the plane. 

The older mean age and spatial age variation we observe in this work is qualitatively similar to that recently found by \citet{Bovy2019}. Their work likely uses similar stars to our work here, but ages are derived in a very different fashion, with different training sets. \citet{Bovy2019} describes the bar as a structure that is distinct in mean age. This is also a prediction of \citet{Ness2014}, who find that the mean age of the off-bar stars in the plane are younger than the mean age of the on-bar stars in the plane. Our maps displayed in Figure \ref{fig:maps} show similar results to \citet{Bovy2019}, in that spatial bins corresponding to the on-bar side of the MW are older than for the off-bar. However, as described in \S \ref{sec:results_onoff}, once we correct for different $|Z_{\rm GC}|$ sampling, we cannot conclude that the on-bar and off-bar populations are different in age at $|Z_{\rm GC}|$ $<$ 0.25 kpc. Future data from sightlines at $b$ = $0^{\circ}$ of the bulge combined with more careful selections (see \citealt{Queiroz2020:inprep}) will more conclusively answer this question.

In addition to a vertical age gradient in the bulge, \citet{Ness2014} find that the youngest stars should be in the plane at $|Z_{\rm GC}|$ $<$ 0.14 kpc. We do find a non-neglible fraction of young stars that share the following properties:

\begin{itemize}
    \item $+$0.2 $<$ [Fe/H] $<$ $+$0.4
    \item 2.0 kpc $\lesssim$ $R_{\rm cy}$ $\lesssim$ 3.0 kpc
    \item $|Z_{\rm GC}|$ $<$ 0.5 kpc
    \item $v_{\phi}$ $\sim$ 150 km/s, or $\sim$ 50 km/s larger than older stars at the same metallicity
\end{itemize}

Therefore, our picture of the bulge from this work is that stars across all metallicities were formed some 8-10 Gyr ago, and can be found all across the bulge. The more planar regions of the bulge enriched quicker than the off-plane regions of the bulge, with the outermost $|Z_{GC}|$ bin exhibiting no stars with [Fe/H] $\gtrapprox$ $+$0.2. However, superposed on this old stellar population is a younger, temporally extended population that appears to be in the plane and at super-solar metallicities. So, after the bulge/bar formed in the initial burst, star formation still occurred, at a lower rate, in a disk or ring, until stopping some 2-4 Gyr ago.

\subsection{Young Star Reconciliation}

The paradigm for many decades was that the bulge only contained old stars, where ``old'' refers to stars with age $\gtrsim$ 8 Gyr. As eloquently summarized in \citet{Nataf2016}, there exist multiple lines of evidence for an exclusively old bulge, as well as multiple lines of evidence for some young-to-intermediate age, metal-rich bulge stars. Because we find our younger stars in the plane to be more metal rich and rotating around the Galactic Center, then this means that studies that are biased against metal-rich stars, look at sight-lines at $b$ $>$ 5$^{\circ}$, or remove stars based on high proper motions, may be removing these younger stars. Given these criteria, we try to reconcile the confirmed presence of younger stars in this work, and other works such as \citet{Bensby2013}, with other studies that find no younger stars.

\citet{Kuijken&Rich2002} reanalyzed the Baade's Window HST data of \citet{Holtzman1993}, whose initial conclusions suggested the super-solar metallicity stars have a median age of $\sim$ 5 Gyr. Baade's Window is at $b$ $\sim$ $-$4$^{\circ}$, so there should be some younger metal-rich stars in these fields. However, in their reanalysis, \citet{Kuijken&Rich2002} found no young stars after cleaning this sample with new proper motion measurements. While a detailed sightline-by-sightline proper motion analysis is beyond the scope of this paper, the young stars we identify would have reasonably large proper motions in the $l$ direction at the longitude of Baade's Window, suggesting a strict proper motion cut could potentially remove these stars.

A similar argument could be made for why young stars are not found in the \citet{Zoccali2003} and \citet{Clarkson2011} samples, although the former sample appears to contain fewer super-solar metallicity stars---likely a consequence of studying a higher latitude field, where the metal-rich stars are a weaker component of the bulge MDF (e.g., Figure \ref{fig:age_feh_relation}, \citealt{Zoccali2017,Garcia-Perez2018}). Future analysis, consisting of a field-by-field proper motion comparison, will determine whether or not the proper motion cleaning is a reason why these photometric studies are lacking in younger bulge stars. However, \citet{Haywood2016} suggest that some of these photometric studies (e.g., \citealt{Valenti2013}), that show a tight main-sequence turnoff, actually necessitate a decent fraction of young, metal-rich stars in the bulge (see also \citealt{Barbuy2018}).

If the works that show an exclusively old bulge can be explained by probing sightlines higher from the plane or proper motion cuts inducing a bias, then the works that show younger stars better fit into the paradigm listed above. The \citet{Bensby2013} sample consists of stars that are nearly all at $|b|$ $<$ 5$^{\circ}$, so they are likely at a $|Z_{\rm GC}|$ height where we see young stars, but we do not know the exact distance to these stars. As already shown in Figure \ref{fig:age_feh_relation}, the younger stars in both of our samples overlap. Younger metal-rich stars are also found by \citet{Bernard2018}, who do use proper-motion cuts, but study stars at  2 $< |b|$ $<$ 4$^{\circ}$. The presence of younger and intermediate-age stars at latitudes closer to the plane is also consistent with the findings of \citet{Catchpole2016}, although they find that the long-period Mira variables (age $\sim$ 5 Gyr) exhibit a clumpy distribution, and suggest that they are associated with the bar.

In summary, the likelihood of observing younger stars in the bulge depends on the metallicity and $|Z_{\rm GC}|$ probed. However, nowhere do we find stars as young as we do in the disk outside of the bulge region (see \S \ref{sec:append_disk}). Therefore, in the areas where we find younger bulge stars, star formation still shut off some $\sim$ 2+ Gyr ago, so any analyses that find statistically significant numbers of stars in the bulge with age $<$ 2 Gyr would still be difficult to reconcile with the present analysis as well as most or all other studies in the literature.

\subsection{The Milky Way in the Galactic Context}

While we would agree that the bulge region of the MW contains numerous old stars, as most studies of external galaxies find, we show that it does depend on the sample location and what metallicities are being probed. \citet{Kruk2018} find that, after decomposing the inner regions of their barred galaxy sample into bar+disk, the disk component is often bluer, or younger. This agrees qualitatively with what we see in the MW. Additionally, \citet{Fragkoudi2020} find that some of the galaxies studied in their barred galaxy sample from Auriga simulations exhibit ``inner rings'' of recent star formation, resulting in a non-negligible fraction of stars with ages $<$ 5 Gyr.

There have also been many studies to try to understand how the barred regions of external galaxies differ from the disk regions in both age and metallicity gradients. Some works show that the gradients along bars are shallower than gradients off the bar (e.g., \citealt{Fraser-McKelvie2019,Neumann2020}), while other works do not show this (e.g., \citealt{SanchezBlazquez2014}). See \cite{Seidel2016} for a more complete description of these discrepancies. A detailed age-gradient study is beyond the scope of this work, but we do find some evidence suggesting a vertical age gradient in at least the off-bar region of the MW, and possibly in the on-bar region as well (see Figure \ref{fig:maps} and \S \ref{sec:results_onoff}). Perhaps this can resolve the tension on whether or not one observes radial age-gradient variations for on-bar and off-bar populations in external galaxies, as the result will depend on distance from the plane probed.

Finally, the fact that we observe slower metallicity evolution with increasing height suggests that, while a bar may be efficient at washing out radial gradients, there still exists some $|Z_{\rm GC}|$ gradient in the SFH that has remained measureable at present times.

\subsection{Primordial Carbon, Nitrogen, and Helium Abundance Variations}
\label{sec:prim_var}

Although we do not explicitly map [C/N] to ages in this work, {\it The Cannon} mainly cues off of C and N spectral features to derive the ages of stars (see e.g., \citealt{Ness2016b}). The ages we derive then implicitly rely on the fact that C and N are dredged up and mixed in similar ways for the bulge stars as they are for stars near the solar circle, which are the stars that comprise our training set.

Therefore, there are three potential caveats of this study that could be affecting our age determination:

\begin{enumerate}
    \item The birth abundance of C and N for the bulge stars differs from the birth abundance of C and N for the solar-circle stars, ultimately resulting in a different post-first dredge-up [C/N] abundance.
    \item The birth abundance helium affects the way C and N are dredged up.
    \item The birth abundance helium itself varies enough to affect the ages of the stars in a drastic way which we are not taking into account.
\end{enumerate}

It is most important to consider the effect these caveats would have on our identification of younger stars in the bulge.

\subsubsection{Birth Abundance C and N Variation}
\label{sec:disc_birth}
It is possible that the mass-dependent dredge-up that allows for the mapping of [C/N] to mass (and then age) may be affected by large differences in the birth C and N abundances, or even more specifically, the birth [C/N] abundance ratio. \citet{Martig2016a} used a sample of APOGEE subgiants, which span no more than 2-3 kpc from the Solar Circle, to show that the birth [C/N] abundance ratio didn't vary wildly with Galactic position. Unfortunately, no large sample of subgiants with precise C and N abundances exist for the bulge.

Despite this, one argument we can make for minimal birth [C/N] abundance variation relative to our  training set [C/N] abundances is that the bulge chemical abundances appear to be very similar to the chemical abundances of the high-$\alpha$ or thick disk stars that are found within the solar circle. These stars appear to share very similar chemical abundance tracks in the $\alpha$-elements, light odd-Z elements, and Fe-peak elements like Ni and Mn (e.g., \citealt{Rojas-Arriagada2017,Zasowski2019,Queiroz2020:inprep}). It would be difficult to form stars that have the same $\alpha$-element abundances, but vastly different C abundances, or stars that have similar metallicity-dependent odd-Z element abundance patterns, but vastly different N abundance patterns. Therefore, because the bulge sample is comprised of stars with $\alpha$-element abundances and metallicities that are represented in our training set, we have no reason to think that the birth abundance [C/N] varies in a way that would cause erroneous ages.

\subsubsection{Helium Affecting Dredge-up}

There is some work in the literature that suggests that helium can affect the dredge-up in giant stars. \citet{Karakas2014} find that, during third dredge-up, an AGB star that is enhanced in helium will dredge up less carbon. However, we are most concerned about post-first dredge-up, as the majority of our stars should be RGB stars given the log(g) range studied. \citet{Salaris2015} found little to no effect of helium abundance on the [C/N] abundance after first dredge-up, but they only considered small changes in helium ($\Delta Y \sim $0.02).  There is some evidence that the bulge contains helium-enhanced stellar populations (e.g., \citealt{Nataf2013,Buell2013}), so this work should be repeated with $\Delta Y$ values of $\sim $ 0.06.

\begin{figure*}[t]
\begin{center}
\includegraphics[width=0.95\hsize,angle=0]{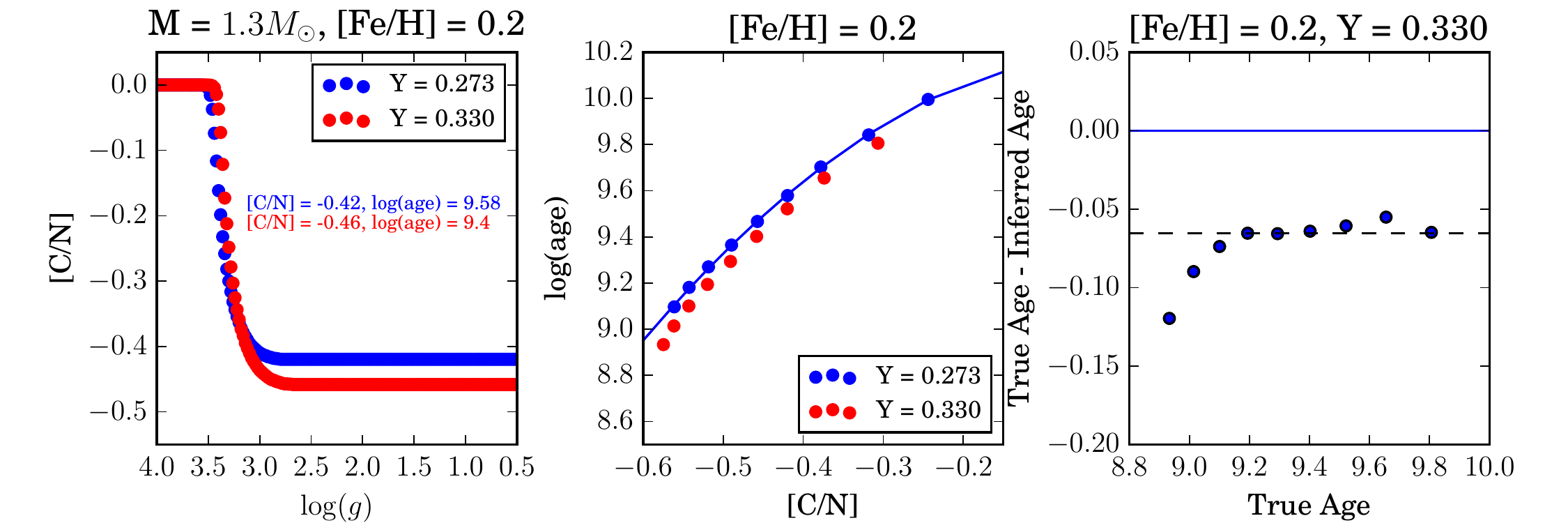}
\caption{Left: [C/N] vs. $\log(g)$ for a 1.3 M$_{\odot}$, [Fe/H] = 0.2 stellar model from \citet{Tayar2017}. Blue is the solar-helium model and red is the helium-enhanced model. Middle: log(age)-[C/N] relation for several solar-helium models (blue) and helium-enhanced models (red). The blue line shows a quadratic fit to the solar-helium log(age)-[C/N] relation. Right: the different between true ages from the helium-enhanced models and ages inferred from the [C/N] abundance ratios after first dredge up for the helium-enhanced models using the relation derived from the solar-helium models in the middle panel.} 
\label{fig:he}
\end{center}
\end{figure*}

To investigate the potential effects of helium abundance on [C/N] abundance after first dredge up, we analyze stellar models from \citet{Tayar2017}. Results are shown in Figure \ref{fig:he}.

In the left panel of Figure \ref{fig:he} we show how the [C/N] abundance ratio changes as a 1.3 M$_{\odot}$, [Fe/H] = 0.2 star ascends the red giant branch for two different helium birth abundances. The star that is enhanced in helium ends up with a lower [C/N] abundance ratio after first dredge up. This means \textit{The Cannon} might assign a younger age than is warranted. However, as shown in the text in the left panel of Figure \ref{fig:he}, the star itself has a younger age because it is helium-enhanced. 

To investigate which effect wins out, we show the ages of stars across a range of masses at [Fe/H] = 0.2 in the middle panel of Figure \ref{fig:he} for solar-helium (blue) and enhanced helium (red) as a function of [C/N] post-first dredge up. Both helium abundances follow a similar log(age)-[C/N] relation, with a slight offset between the two such that at fixed [C/N] the helium-enhanced stars are actually slightly younger. We fit a quadratic function to the solar-helium stars and derive a log(age)-[C/N] relation. We use this relation to infer an age of the helium-enhanced stars based on their [C/N] after first dredge up. The right panel of Figure \ref{fig:he} shows the difference between the true age of these helium-enhanced stars and the ages inferred based on [C/N]. We find that the inferred ages are $\sim$ 0.06 dex older than the true ages. Therefore, our age method would actually put helium-enhanced stars slightly \emph{older} than they actually are.

\subsubsection{Helium Affecting Stellar Age}

Stars enhanced in helium live shorter lifetimes. \citet{Bazan1990} suggest that the lifetime of a star can change by $\sim$ 0.3 dex in {$\rm \log(age)$} for helium Y = 0.3 for lower-mass stars. If the [C/N] abundance ratio after first dredge-up is not significantly impacted by birth helium abundance, then our age method would be pushing younger, helium-enhanced stars to older ages. This would mean that our sample would contain even more younger stars than what we currently find.

After all this, even if the helium conspires to give us younger ages in one way or the other, then instead of young stars, we've simply found ``helium-peculiar'' stars, and they tend to be found at $+$0.2 $<$ [Fe/H] $<$ $+$0.4, low-$|Z_{\rm GC}|$, $R_{\rm cy}$= 2-3 kpc, etc.

\section{Summary}

We summarize our conclusions below:

\begin{enumerate}
    \item We have derived ages for 47,000 luminous giant stars across much of the MW disk and bulge.
    \item The stars at $-$0.5 $<$ [Fe/H] $<$ $+$0.5 that reside in the bulge are primarily old, with about half of the stars in our sample being older than 8 Gyr.
    \item We find that the stars in the plane enriched at a rate of $dZ/dt \sim 0.0034$  ${\rm Gyr}^{-1}$, which is exactly what was found by \citet{Haywood2016}. However, stars out of the plane enriched at a much slower rate ($dZ/dt \sim 0.0013$ ${\rm Gyr}^{-1}$, a factor of three slower), suggesting an inside-out bulge formation scenario where the stellar populations have not yet been fully mixed.
    \item We find a non-negligible fraction of younger stars (2-5 Gyr old) that primarily have $+$0.2 $<$ [Fe/H] $<$ $+$0.4, $|Z_{\rm GC}| < $ 0.25 kpc, 2.0 kpc $\lesssim R_{\rm cy} \lesssim$ 3.5 kpc, and kinematics more consistent with rotation. This suggests an extended star-formation history of the bulge that took place in a disk after the initial burst.
    \item This work suggests that some of the literature disputes regarding an old versus young bulge can be settled by understanding that the younger stars are only found at $|Z_{\rm GC}|$ $\lesssim$ 0.5 kpc and at [Fe/H] $\gtrsim$ 0.2.
    \item When correcting for $|Z_{\rm GC}|$ sampling effects, we do not find a measurable age difference between stars inside of the bar and stars outside of the bar, but future data will improve this measurement, as well as measurements of the vertical and radial gradients.
\end{enumerate}

\acknowledgments
We thank the anonymous referee for their comments that improved this manuscript. We thank Melissa Ness for the extremely helpful guidance on how best to run {\it The Cannon}. We thank Jamie Tayar for the help in providing and explaining her stellar models. 

S.H. is supported by an NSF Astronomy and Astrophysics Postdoctoral Fellowship under award AST-1801940. D.K.F. was supported by the grant 2016-03412 from the Swedish Research Council. DAGH acknowledges support from the State Research Agency (AEI) 
of the Spanish Ministry of Science, Innovation and Universities (MCIU) 
and the European Regional Development Fund (FEDER) under grant 
AYA2017-88254-P.  ARL acknowledges substantial financial support for this research from FONDECYT Regular 1170476 and QUIMAL project 130001. T.C.B. acknowledges partial support from Grant PHY 14-30152 (Physics Frontier Center/JINA-CEE), awarded by the U.S. National Science Foundation. D.G. gratefully acknowledges support from the Chilean Centro de Excelencia en Astrof{\'i}sica
y Tecnolog{\'i}as Afines (CATA) BASAL grant AFB-170002.
D.G. also acknowledges financial support from the Direcci{\'o}n de Investigaci{\'o}n y Desarrollo de
la Universidad de La Serena through the Programa de Incentivo a la Investigaci{\'o}n de
Acad{\'e}micos (PIA-DIDULS). DMN acknowledges support from NASA under award Number 80NSSC19K0589, and support from the  Allan C. And Dorothy H. Davis Fellowship.

Funding for the Sloan Digital Sky Survey IV has been provided by the Alfred P. Sloan Foundation, the U.S. Department of Energy Office of Science, and the Participating Institutions. SDSS-IV acknowledges support and resources from the Center for High-Performance Computing at the University of Utah. The SDSS web site is www.sdss.org.

SDSS-IV is managed by the Astrophysical Research Consortium for the Participating Institutions of the SDSS Collaboration including the Brazilian Participation Group, the Carnegie Institution for Science, Carnegie Mellon University, the Chilean Participation Group, the French Participation Group, Harvard-Smithsonian Center for Astrophysics, Instituto de Astrof\'isica de Canarias, The Johns Hopkins University, Kavli Institute for the Physics and Mathematics of the Universe (IPMU) / University of Tokyo, Lawrence Berkeley National Laboratory, Leibniz Institut f\"ur Astrophysik Potsdam (AIP), Max-Planck-Institut f\"ur Astronomie (MPIA Heidelberg), Max-Planck-Institut f\"ur Astrophysik (MPA Garching), Max-Planck-Institut f\"ur Extraterrestrische Physik (MPE), National Astronomical Observatories of China, New Mexico State University, New York University, University of Notre Dame, Observat\'ario Nacional / MCTI, The Ohio State University, Pennsylvania State University, Shanghai Astronomical Observatory, United Kingdom Participation Group, Universidad Nacional Aut\'onoma de M\'exico, University of Arizona, University of Colorado Boulder, University of Oxford, University of Portsmouth, University of Utah, University of Virginia, University of Washington, University of Wisconsin, 
Vanderbilt University, and Yale University.

This research made use of Astropy \footnote{http://www.astropy.org} a community-developed core Python package for Astronomy \citep{astropy:2013,astropy:2018}, SciPy \citep*{SciPy}, NumPy \citep{NumPy}, and Matplotlib \citep{Hunter:2007}.

\bibliographystyle{apj}
\bibliography{ref_og.bib}

\appendix
\section{Age Validation}
\label{sec:append_age_verify}
\subsection{North vs. South}
\label{app:nvs}
While tests of the two APOGEE spectrographs suggest that their performance is nearly identical \citep{Wilson2019}, there are small variations in the line spread function (LSF) across the detectors in both instruments, as well as variations in the LSF between the two instruments. Therefore, we might expect some differences in our results for Northern spectra than for Southern spectra, especially since the training set is $\sim$ 75\% Northern spectra. Because several stars we derive ages for were observed from both the Northern and Southern instrument setups, we are able to quantify potential systematic uncertainties in age derivation based on whether the star is observed from the Northern or Southern Hemisphere. 

\begin{figure*}[t]
\includegraphics[width=1.0\hsize,angle=0]{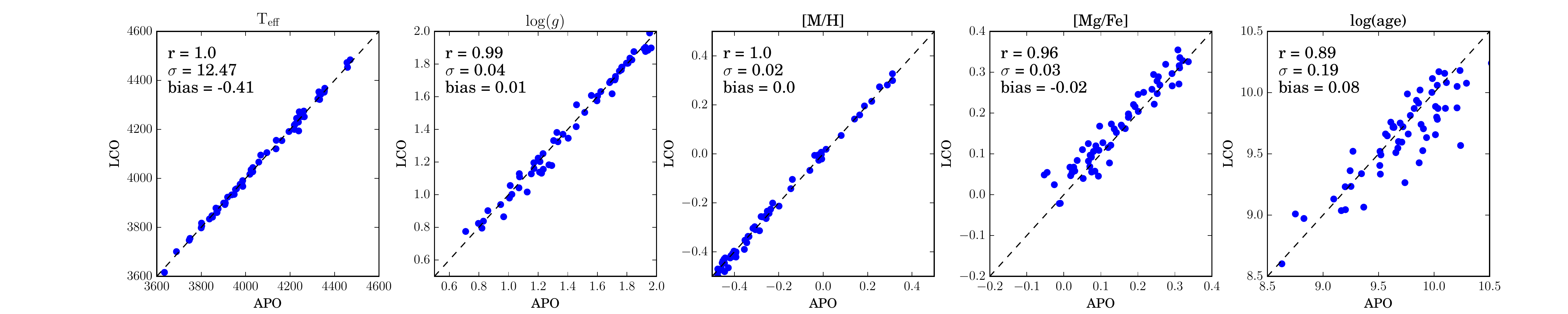}
\caption{Comparison of labels derived for 62 stars observed from both the northern and southern hemisphere instruments. Stars were selected to have S/N $>$ 70 in both observations. Correlation coefficients, standard deviations, and biases are indicated in the upper-left of each panel.}
\label{fig:north_v_south}
\end{figure*}

These  comparisons are shown in Figure \ref{fig:north_v_south}. Only those stars with S/N $>$ 70 from both hemispheres are plotted. The ${\rm T}_{\rm eff}$, $\log(g)$, and [M/H] labels output by {\it the Cannon} are nearly identical. There may be a slight bias in derived [Mg/Fe], such that the APO [Mg/Fe] values are 0.02 dex lower than the LCO values. The ASPCAP [Mg/Fe] values for the same set of stars differ by 0.01 dex such that the APO values again are $\sim$ 0.01 dex lower than the LCO  values. For the analysis in this paper, we use the ASPCAP [Mg/Fe] values, but note that we only use these values to bin stars into mono-abundance bins, and adopting either set of [Mg/Fe] values has no significant effect on our results.

There is also a slight bias in {$\rm \log(age)$}, such that the APO ages are 0.08 dex older than the LCO ages, on average. In our analysis, we apply a 0.08 dex offset to the LCO ages to bring this overlap sample into better agreement. This offset is included in the ages provided in Table \ref{tab:ages}. This offset does not significantly influence any of our results.

\subsection{Deriving Uncertainties}
\label{sec:append_uncert}

The 90-10 cross-validation test done in $\S$ \ref{sec:training} suggests an age precision of $\sim$ 0.22 dex (standard deviation of the differences, 0.31, divided by $\sqrt{2}$). However, we find that this precision is not uniform across parameter space. In particular, the precision varies strongly with $\log(g)$. We approximate this variation by calculating the standard deviation of the differences between the input age label and the label obtained during the 90-10 cross-validation step. We calculate this for 20 stars bins in a ``moving boxcar'' fashion, and fit a function to these standard deviations as a function of $\log(g)$. The distribution of differences between two samples is related to the uncertainty of a single sample by $\sqrt{2}$, so these standard deviations are divided by $\sqrt{2}$. We fit the following function to these results, and this is the prescription for the uncertainties reported in Table \ref{tab:ages}.

$$\sigma_{\log({\rm age})} = 0.42-0.19\log(g)+0.035(\log(g))^{2}$$

\subsection{Comparison to Open Clusters}
\label{sec:append_cluster}

We match our age sample to the \emph{Gaia} cluster catalog provided by \citet{Cantat-Gaudin2018}. We find 3 clusters for which we have derived ages for $>$ 3 stars: NGC 6791 (9 stars), NGC 6819 (6 stars), and NGC 2204 (6 stars). We only select stars that have \citet{Cantat-Gaudin2018} membership probabilities of $>$ 0.5. We compare the median age we derive for the stars in these clusters to the ages provided in the \citet{Kharchenko2013} cluster catalog. The log(age) values we adopt from this catalog are 9.65 for NGC 6791, 9.21 $\pm$ 0.024 for NGC 6819, and 9.29 $\pm$ 0.018 for NGC 2204. The log(age) values we derive in this work, including the dispersion in log(age) for these clusters are 9.86 $\pm$ 0.23 for NGC 6791, 9.11 $\pm$ 0.12 for NGC 6819, and 9.13 $\pm$ 0.07 for NGC 2204. 

We find that we slightly underestimate the ages for the youngest two clusters by $\sim$ 0.1 dex, but overestimate the age of the oldest cluster, NGC 6791, by $\sim$ 0.2 dex. However, some literature work suggests that NGC 6791 could be as old as log(age) $\sim$ 9.9 if helium enhancement is accounted for (e.g., \citealt{Brogaard2012}), which would agree much better with the age we would find for NGC 6791 using our 9 members.

\subsection{Comparison to APOKASC}
\label{sec:append_apokasc}

\begin{figure}[t]
\includegraphics[width=1.0\hsize,angle=0]{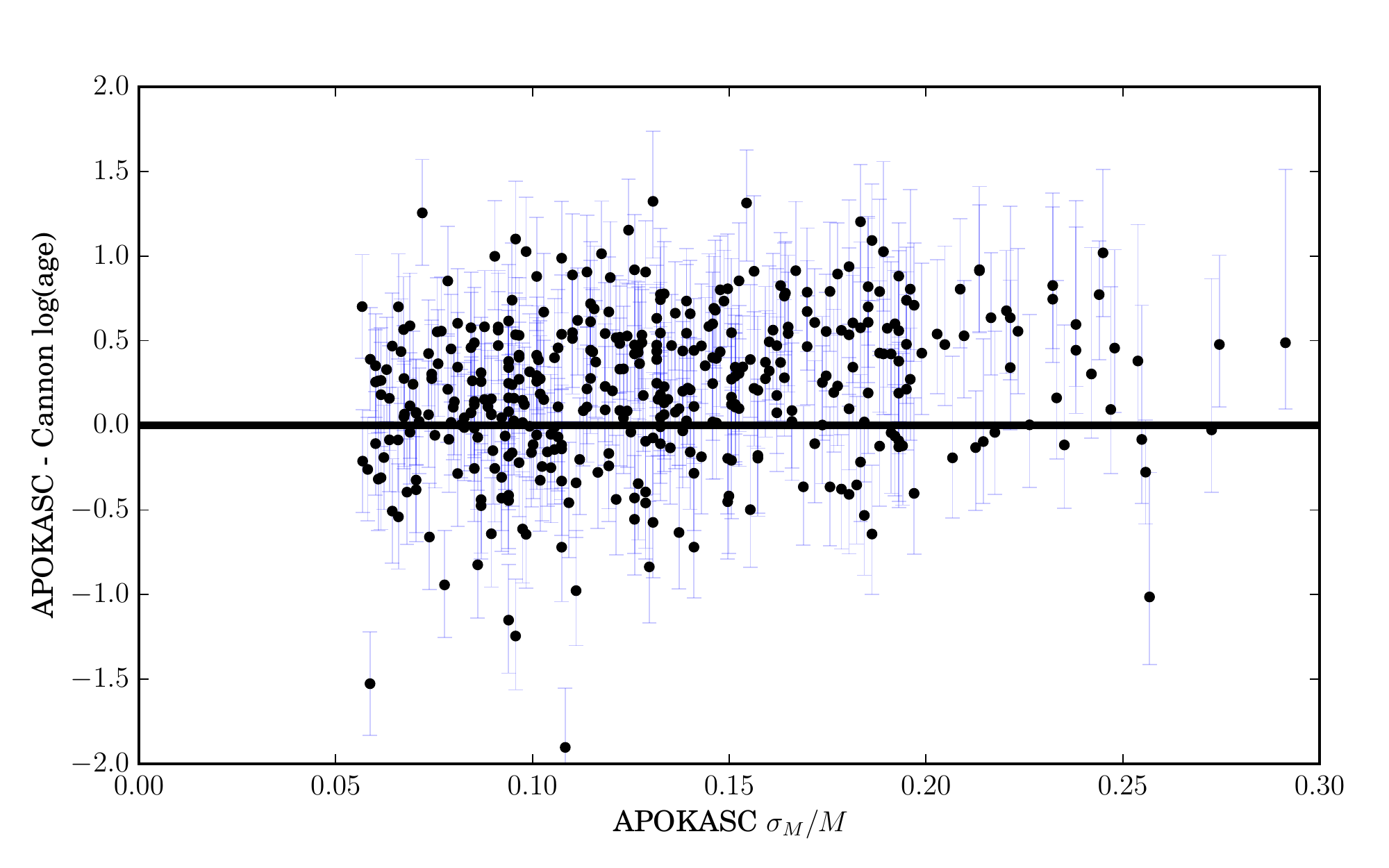}
\caption{Differences between APOKASC-2 {$\rm \log(age)$} and {\it The Cannon} {$\rm \log(age)$} ages derived in this work, plotted as a function of fractional APOKASC-2 mass uncertainty.}
\label{fig:apokasc_comp}
\end{figure}

We compare our ages to those derived from asteroseismic masses in the APOKASC-2 sample \citep{Pinsonneault2018}. This overlap sample is relatively small (412 stars), and only contains stars with $\log(g)$ between 1.5 and 2.0, but it serves as one of the only external checks we have for accuracy. Figure \ref{fig:apokasc_comp} shows that the ages agree well within the scatter for stars with APOKASC-2 mass uncertainty $<$ 10\%, and there is no apparent systematic offset. However, above an APOKASC-2 mass uncertainty of 10\%, there is an offset such that the APOKASC-2 ages are on average older than {\it The Cannon} ages. A detailed understanding of this apparent offset with uncertainty is beyond the scope of this work, but other studies that have not removed APOKASC-2 stars with large mass uncertainties have found that the ages they reproduce are generally younger than the input APOKASC-2 ages at older ages (e.g., \citealt{Martig2016a,Mackereth2019b}), and these works typically apply their own correction factors.

We do not apply a correction to bring our results into better agreement with the APOKASC ages because it is the APOKASC stars with large mass uncertainties that are discrepant, and the vast majority of stars studied in this paper have $\log(g)$ $<$ 1.5, so this potential offset is only motivated for a small fraction of our sample.

\subsection{Reproducing MW Disk Results}
\label{sec:append_disk}

\begin{figure}[t]
\includegraphics[width=1.0\hsize,angle=0]{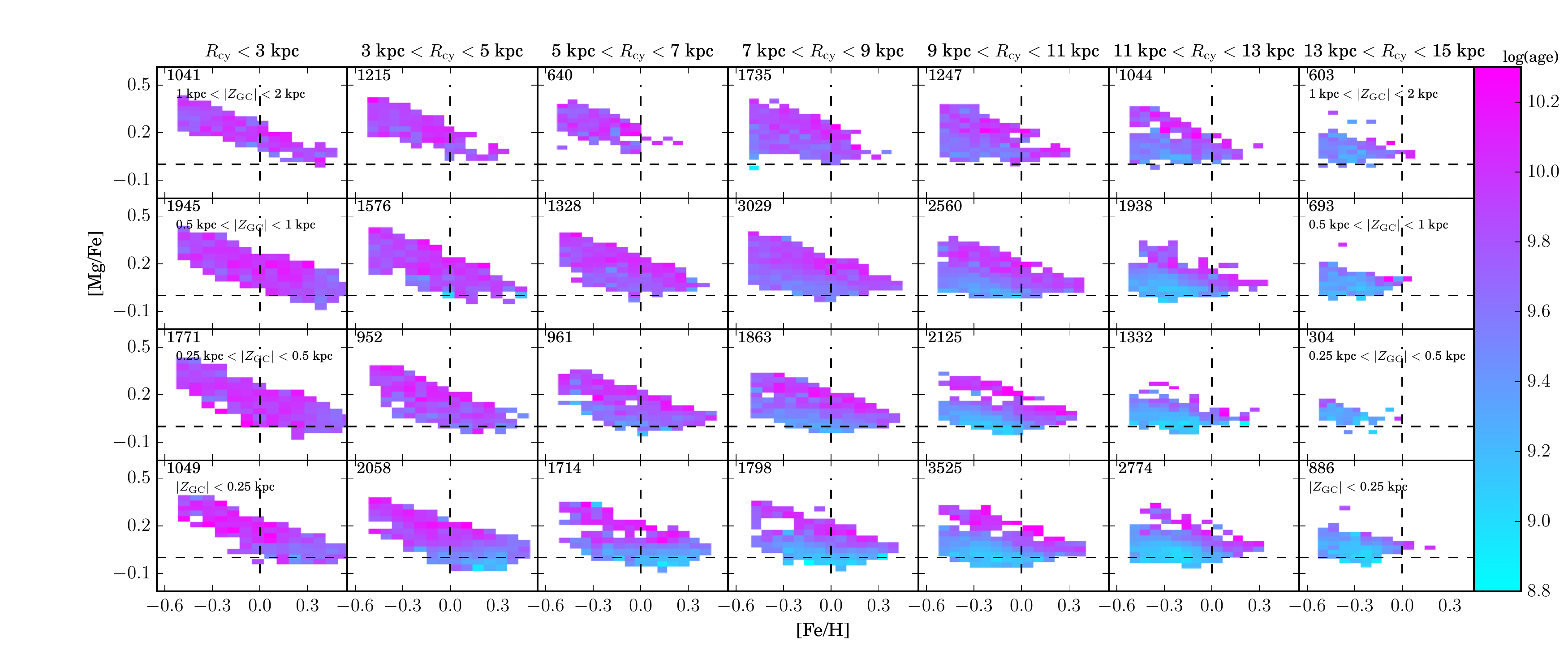}
\caption{[Mg/Fe]-[Fe/H] maps of the MW colored by log(age). Columns are separate by bins of $R_{\rm cy}$ and rows are separated by bins of $Z_{\rm GC}$.}
\label{fig:disk}
\end{figure}

To show further support for the accuracy of the ages derived in this study, we reproduce ages maps of the MW that have been produced in other works. Our sample is different in that the ages are derived using a different training set than these other works, and we study the more luminous giants, whereas these other works typically study red clump giants and/or giants farther down the giant branch. These maps are shown in Figure \ref{fig:disk}. As mentioned in \S \ref{sec:results}, the youngest stars in our full sample are not found in the bulge, but at $R_{\rm cy} >$ 3 kpc and $|Z_{\rm GC}| <$ 0.25 kpc. As has been found in other studies, the youngest stars in each radial bin in the planar region of the Galaxy (bottom row of Figure \ref{fig:disk}) are found at lower [Fe/H] from the inner to outer Galaxy (e.g., \citealt{Ness2016b,Martig2016b,Mackereth2019b}). We also select stars in the same Galactic region as the CoRoGEE sample and measure a metallicity gradient of stars with age $<$ 2 Gyr and a gradient of stars with age $>$ 10 Gyr. We find these gradients to be $-0.068 \pm 0.001$ dex/kpc and $-0.030 \pm 0.003$ dex/kpc, respectively, which are both in excellent agreement with the same gradients measured by \citet{Anders2017b}.

We also see that stars increase in age from near the plane to above the plane, but the stars above the plane themselves exhibit a radial age gradient, qualitatively similar to what was found by \citet{Martig2016b}. 

\subsection{LMC and Sgr ages}
\label{sec:append_dwarfs}

APOGEE has also observed stars in the Large Magellanic Cloud (LMC) and Sagittarius dwarf galaxy (Sgr), allowing us to further validate our ages. Because we are limited to deriving ages for stars with [Fe/H] $>$ -0.5, we can only look at the ages for the most metal-rich stars in these two galaxies. The most metal-rich stars of these galaxies have [Fe/H] $\sim$ 0.0, so the following analysis is limited to stars with -0.5 $<$ [Fe/H] $<$ 0.0.

\begin{figure}[t]
\includegraphics[width=0.5\hsize,angle=0]{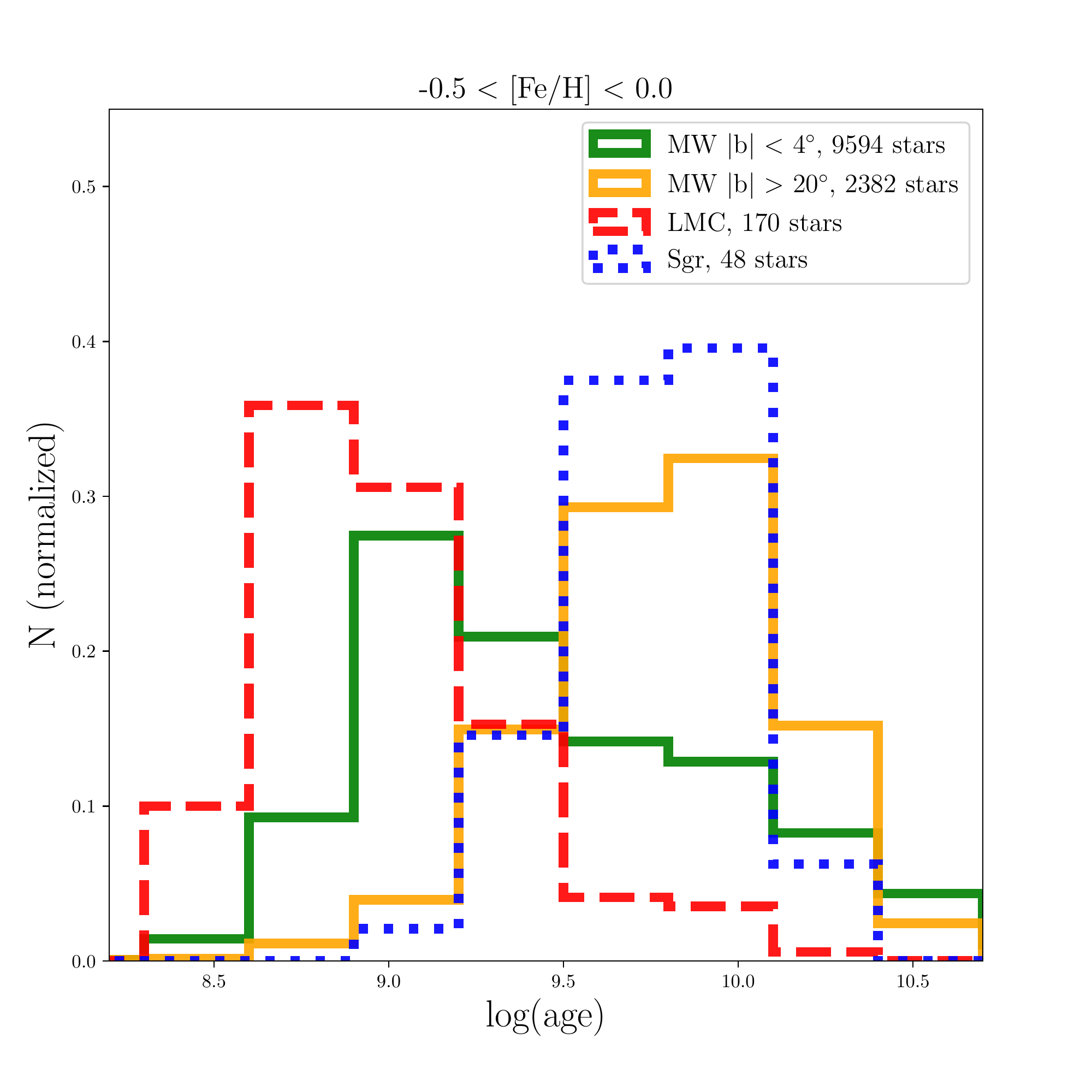}
\caption{Age distributions of the four samples described in the text, limited to -0.5 $<$ [Fe/H] $<$ 0.0. }
\label{fig:dwarfs}
\end{figure}

In Figure \ref{fig:dwarfs} we compare the age distributions of the two dwarf galaxies (LMC in red and Sgr in blue) to the age distribution of a low-latitude MW disk sample (green) and a high-latitude MW disk sample (orange). LMC stars are selected from the \citet{Nidever2020} sample and Sgr stars are selected from the \citet{Hayes2020} sample. The MW stars are split in latitude to select a ``thin disk'' sample ($|$b$|$ $<$ $4^{\circ}$) and a ``thick disk'' sample ($|$b$|$ $>$ $20^{\circ}$).

We find that the LMC stars are young, with a median age of $\sim$ 0.8 Gyr. This is consistent with the expected age of these most metal-rich LMC stars based on the star formation history of \citet{Harris&Zaritsky2009} as well as the the chemical evolution model invoked in \citet{Nidever2020} to explain the observed [Mg/Fe]-[Fe/H] abundance pattern. The Sgr stars are older, with a median age of $\sim$ 6 Gyr. This is consistent with N-body simulations of Sgr, which generally find that Sgr fell into the MW some 5-6 Gyr ago (e.g., \citealt{Law&Majewski2010}). Moreover, the SFH from \citet{Siegel2007} suggests the bulk of Sgr stars at these metallicities should be 4-8 Gyr old. We also find that, as expected, the thin disk MW stars are younger than the thick disk MW stars ($\sim$ 2 Gyr old as compared to $\sim$ 6 Gyr old).

\end{document}